\documentclass[11pt]{article} 

\newif\ifFULL
\FULLfalse    

\usepackage[margin=1in,includefoot=false,footskip=30pt]{geometry}
\usepackage[font=footnotesize]{caption}
\usepackage{comment}
\usepackage{amsmath}
\usepackage{amsthm}
\usepackage{amssymb}
\usepackage[normalem]{ulem}
\usepackage{url}
\usepackage[final]{showkeys}
\usepackage{tikz}
\usetikzlibrary{matrix, positioning, fit, shapes.geometric, calc, arrows.meta}
\usepackage{lineno}
\usepackage{thmtools}
\usepackage{xcolor}
\definecolor{mydeepred}{HTML}{c51b1d}
\usepackage[colorlinks=true, allcolors=mydeepred, hypertexnames=false]{hyperref}
\usepackage[nosort]{cleveref}
\usepackage{multirow}
\usepackage{booktabs}
\usepackage{refcount}
\usepackage{macros}

\newtheorem{theorem}{Theorem}
\newtheorem{lemma}[theorem]{Lemma}

\newtheorem{claim}[theorem]{Claim}
\newtheorem{corollary}[theorem]{Corollary}

\newtheorem*{hypothesis*}{Hypothesis}
\newtheorem*{conjecture*}{Conjecture}
\newtheorem*{claim*}{Claim}
\theoremstyle{remark}
\newtheorem{remark}[theorem]{Remark}
\theoremstyle{definition}
\newtheorem{definition}[theorem]{Definition}
\newtheorem*{definition*}{Definition}

\begin{document}

\title{Hardness of Regular Expression Matching with Extensions}
\author{
    Taisei Nogami\\ Waseda University \\ \texttt{\small sora410@fuji.waseda.jp}
    \and Yoshiki Nakamura\\ Chiba University \\ \texttt{\small nakamura.yoshiki.ny@gmail.com}
    \and Tachio Terauchi\\ Waseda University \\ \texttt{\small terauchi@waseda.jp}
}
\date{}
\begin{titlepage}
\maketitle
\thispagestyle{empty}
\begin{abstract}
    The regular expression matching problem asks whether a given regular expression of length $m$ matches a given string of length $n$, and its time complexity is fairly well understood.
    Meanwhile, regular expressions have been extended to support various extensions for both theoretical and practical reasons, which can substantially change the time complexity of the matching problem.
    In this work, we consider four well-known extensions to regular expressions called \emph{intersection}, \emph{squaring}, \emph{complement} and \emph{backreference}, and we prove a number of novel conditional time complexity lower bounds for regular expression matching with these extensions, assuming the Orthogonal Vectors Conjecture (OVC), the $k$-Orthogonal Vectors Conjecture ($k$-OVC) or the $k$-Clique hypotheses:
\begin{itemize}
    \item For each of intersection and squaring (for squaring, even without Kleene star), no algorithm runs in $n^{2-\varepsilon} \poly(m)$ time under moderate-dimension OVC, $n^{2-\varepsilon} 2^{\asympo(m)}$ time under log-dimension OVC or $n^{\omega-\varepsilon} 2^{\asympo(m)}$ time ($2 \le \omega < 2.3714$ is the exponent of square matrix multiplication) under the $k$-Clique Hypothesis, and no combinatorial algorithm runs in $n^{3-\varepsilon} 2^{\asympo(m)}$ time under the Combinatorial $k$-Clique Hypothesis, for any constant $\varepsilon > 0$.
    \item For complement (even without Kleene star), no algorithm runs in $n^{2-\varepsilon} \tower(\asympo(m/\log{m}))$ time under moderate-dimension OVC or $n^{\omega-\varepsilon} \tower(\asympo(m/\log{m}))$ time under the $k$-Clique Hypothesis, and no combinatorial algorithm runs in $n^{3-\varepsilon} \tower(\asympo(m/\log{m}))$ time under the Combinatorial $k$-Clique Hypothesis, for any constant $\varepsilon > 0$.
        Furthermore, we resolve the inconsistency between our results and the $\asympO(n^2 \cdot (\log n + m) \cdot 2^m)$-time combinatorial algorithm of Rosu (FoSSaCS'07) by identifying an issue with their algorithm.
    \item For backreference, no algorithm for expressions with at least $k$ capturing groups runs in $n^{2k-\varepsilon} \poly(m)$ time under moderate-dimension $2k$-OVC or $n^{2k-\varepsilon} 2^{\asympo(\sqrt{m})}$ time under log-dimension $2k$-OVC, for any constants $\varepsilon > 0$ and $k \ge 1$.
        Moreover, for expressions of a straight-line form with a single capturing group, no algorithm runs in $n^{2-\varepsilon} \poly(m)$ time under moderate-dimension OVC or $n^{2-\varepsilon} 2^{\asympo(\sqrt{m})}$ time under log-dimension OVC.
\end{itemize}

Our hardness results are tight in the following sense.  They show that for intersection, squaring, complement and a certain subclass of backreference, the known upper bounds for the problems are (conditionally) optimal in that the dependence on $n$ cannot be reduced by any polynomial factor while keeping the dependence on $m$ polynomial (or even elementary, for complement).
Notably, for intersection and complement, our results show that Hopcroft and Ullman's approach from 1979 was already optimal in this sense, and shed light on why the theoretical computer science community has struggled to improve the time complexity of these problems with respect to $n$ and $m$ for more than 45 years.

Additionally, to obtain our results for intersection, squaring and complement in a unified manner, we introduce novel problems called \emph{Dyck Selection}, \emph{Orthogonal Dyck Selection} and \emph{Generalized Dyck Selection}.  They are arguably natural formal language and stringology problems that are based on Greibach's hardest context-free language, a fundamental concept in formal language theory, and capture the essence of why the regular-expression matching problems with these extensions are hard.  Also, as a byproduct, this approach establishes that the matching problem with squaring is LOGCFL-complete.
\end{abstract}

\end{titlepage}

\section{Introduction}
\label{sec:intro}

Regular expressions, originally introduced by Kleene~\cite{kleene1956}, are one of the most successful concepts in computer science, both in theory and practice. 
Among problems related to regular expressions, the \emph{regular expression matching problem} is of foundational importance.  It asks whether a given regular expression $r$ of length $m$ matches a given string $w$ of length $n$.  
As is well known, the problem can be solved in $\asympO(nm)$ time using an algorithm introduced by Thompson~\cite{thompson1968programming}.  Roughly, the algorithm converts a given regular expression $r$ to an equivalent nondeterministic finite automaton (NFA) and simulates it on a given string $w$. 
On the hardness side, Backurs and Indyk showed that the problem cannot be solved in $\asympO((nm)^{1-\varepsilon})$ time for any $\varepsilon > 0$ under the Strong Exponential Time Hypothesis (SETH)~\cite{backurs2016regular}, followed by further developments~\cite{DBLP:conf/focs/BringmannGL17,DBLP:conf/icalp/AbboudB18,DBLP:conf/esa/Schepper20}.  Therefore, the time complexity of regular expression matching is fairly well understood.

Meanwhile, regular expression matching becomes much harder when \emph{extensions} are present.  
Prominent examples include \emph{intersection} $r_1 \regand r_2$ and \emph{complement} $\regnot r$, natural extensions from a set-theoretic point of view, which allow taking intersections and complements of expressions.
Note that complement is at least as expressive as intersection because intersection can be expressed using complement and alternation with only a linear increase in the length of the expression.
%
In 1979, Hopcroft and Ullman showed in their textbook~\cite[Exercise 3.23]{hopcroft1979book} that regular expression matching extended with complement can be solved in $\asympO(n^3 m)$ time.  For many years thereafter, no polynomial improvement over this with respect to $n$ and $m$ had been obtained until the work of Bille, G{\o}rtz and Jessen~\cite{DBLP:journals/corr/abs-2510-09311}.  \cite{DBLP:journals/corr/abs-2510-09311} made an important observation that Hopcroft and Ullman's algorithm can run in $\asympO(n^\omega m)$ time using fast matrix multiplication without significantly altering the algorithm's flow, where $2 \le \omega < 2.3714$~\cite{DBLP:conf/soda/AlmanDWXXZ25} is the exponent of square matrix multiplication.
In addition, it follows from \cite{DBLP:journals/corr/abs-2510-09311} that the algorithm can be extended without changing its running time to handle another extension called \emph{squaring} $\regsq{r}$~\cite{DBLP:conf/focs/MeyerS72}, which denotes the concatenation of $r$ with itself (without doubling the length of $r$).

Another well-known extension is \emph{backreference}, an extension that allows reusing previously matched substrings.  The extension is practically popular and supported by the regular expression engines in the standard libraries of many popular programming languages including Java, Python, JavaScript and more.
The matching problem extended with backreference is NP-hard~\cite{aho1991bref}, and the problem for expressions restricted to at most $\capn$ capturing groups can be solved in $\asympO(n^{2\capn+1} m)$ time~\cite{schmid2016characterising,DBLP:journals/jalc/Schmid24,DBLP:conf/mfcs/NogamiT25}. 
Recently, Uezato showed an $\asympO(n^2 m^2)$-time matching algorithm for expressions of a straight-line form with a single capturing group~\cite{DBLP:conf/cpm/Uezato26}.

In light of the above, we pose the following question: \emph{How tight are these upper bounds?}
Notably, recent studies have shown a surprising fact that the matching problem extended with \emph{lookaround}, a practically popular extension often used to mimic intersection and complement, can actually be solved in $\asympO(nm)$ time~\cite{mamouras2024efficient,barriere2024linear}, and therefore, we have tight bounds for regular expression matching without extensions as mentioned in the first paragraph as well as for that extended with lookaround.  However, tight bounds have not yet been established for the extensions mentioned above.  Namely, for intersection or complement, the strongest negative result derivable from the known results is that neither extension admits an $\asympO((n+m)^{\omega-\varepsilon})$-time algorithm nor a combinatorial $\asympO((n+m)^{3-\varepsilon})$-time algorithm under the $k$-Clique hypotheses, as will be discussed later in \cref{rem:chain} of \cref{sec:intro-res-ere}.
    As for backreference, in a recent independent work, Kumabe and Uezato~\cite{DBLP:conf/icalp/KumabeU26} showed that matching with at least $k$ capturing groups admits no $\asympO(n^{2k-\varepsilon})$-time algorithm under (moderate-dimension) $2k$-Orthogonal Vectors Conjecture for every constant $k \ge 1$, but by using a non-standard form of backreference that is unsupported in most actual regular expression engines; see Section~\ref{sec:intro-res-rewb} for details.

\subsection{Our Results}
\label{sec:intro-res}

In this paper, we consider the four extensions described above, intersection, squaring, complement and backreference, and establish novel conditional lower bounds for regular expression matching with each extension assuming the Orthogonal Vectors Conjecture (OVC), the $k$-Orthogonal Vectors Conjecture ($k$-OVC) or the $k$-Clique hypotheses.
Our results significantly close the gaps between the bounds mentioned above, with some of our lower bounds even being (conditionally) optimal.
Note that our lower bounds under OVC or $k$-OVC also hold under SETH~\cite{DBLP:journals/jcss/ImpagliazzoP01} because SETH implies OVC and $k$-OVC~\cite{DBLP:journals/tcs/Williams05,DBLP:conf/soda/WilliamsY14,DBLP:journals/corr/abs-2204-11681}.

\subsubsection{Hardness of Matching with Intersection and Squaring}
\label{sec:intro-res-semiere-rsq}

We refer to regular expressions extended with intersection, complement and squaring as \emph{semi-extended regular expressions (semi-EREs)}, \emph{extended regular expressions (EREs)} and \emph{RSQs}, respectively.  \emph{Star-free} means containing no Kleene star.

\begin{theorem} \label{thm:semiere-rsq}
    Let $\varepsilon > 0$ be an arbitrary constant.  For semi-ERE matching and RSQ matching (even for star-free RSQ matching), the following hold:
    \begin{enumerate}
    \renewcommand{\theenumi}{(\alph{enumi})}
    \renewcommand{\labelenumi}{\textup{\theenumi}}
        \item \label{thm:semiere-rsq-ov} Unless moderate-dimension OVC fails, no algorithm runs in $n^{2-\varepsilon}\poly(m)$ time (with log-dimension OVC, $n^{2-\varepsilon} 2^{\asympo(m)}$ time).
        \item \label{thm:semiere-rsq-clique} Unless the $k$-Clique Hypothesis fails, no algorithm runs in $n^{\omega-\varepsilon} 2^{\asympo(m)}$ time.  Moreover, unless the Combinatorial $k$-Clique Hypothesis fails, no combinatorial algorithm runs in $n^{3-\varepsilon} 2^{\asympo(m)}$ time.
    \end{enumerate}
\end{theorem}

\cref{thm:semiere-rsq}\ref{thm:semiere-rsq-clique} implies that the $\asympO(n^3 m)$-time combinatorial ERE matching algorithm of Hopcroft and Ullman~\cite{hopcroft1979book} is (conditionally) optimal among combinatorial semi-ERE matching algorithms in the sense that, assuming the Combinatorial $k$-Clique Hypothesis, the exponent of $n$ cannot be shaved by any constant while preserving the factor from $m$ to be subexponential.
Similarly, the $\asympO(n^\omega m)$-time algorithm using fast matrix multiplication observed by Bille, G{\o}rtz and Jessen~\cite{DBLP:journals/corr/abs-2510-09311} is optimal in this sense assuming the $k$-Clique Hypothesis.
We note that if matrix multiplication can be done in quadratic time (i.e.,~$\omega = 2$), \cref{thm:semiere-rsq}\ref{thm:semiere-rsq-ov} also establishes such optimality of the algorithm using fast matrix multiplication under log-dimension OVC.

Moreover, \cref{thm:semiere-rsq} formally captures the threshold in the trade-off between $n$ and $m$ because the exponent of $n$ drops sharply to $1$ once an exponential dependence on $m$ is allowed. 
Indeed, by a standard construction, a regular expression extended with either intersection or squaring (or both) can be converted into an equivalent NFA of $2^{\asympO(m)}$ states.  Combining this with Thompson's algorithm mentioned earlier gives an $n2^{\asympO(m)}$-time matching algorithm for these expressions.
Thus, \cref{thm:semiere-rsq} gives a tight lower bound in that it establishes the conditionally optimal exponents of $n$ in running times of the form $n^c \cdot g(m)$ for both semi-ERE matching and RSQ matching based on whether $g$ is exponential or not.

Additionally, we complement the results of Petersen~\cite{DBLP:conf/stacs/Petersen02}, who showed that semi-ERE matching is LOGCFL-hard and regular expression matching extended with both intersection and squaring is in LOGCFL, by showing that star-free RSQ matching is LOGCFL-hard.  Hence, semi-ERE matching and (star-free) RSQ matching are both LOGCFL-complete.
We note that (star-free) ERE matching is known to be P-complete~\cite{conf/dcagrs/Petersen00}.
\begin{theorem} \label{thm:rsq-logcfl}
    Star-free RSQ matching is LOGCFL-hard.
\end{theorem}

Taken together, these results show that semi-ERE matching and RSQ matching have very similar computational complexity.

\subsubsection{Hardness of Matching with Complement}
\label{sec:intro-res-ere}

As mentioned earlier, EREs subsume semi-EREs.  Hence, the conditional lower bounds in \cref{thm:semiere-rsq} also apply to ERE matching.  However, we can establish much stronger conditional lower bounds for ERE matching in terms of the dependence on $m$.  Let
\[
    \tower(m) = \twotower{m}\,,
\]
i.e., the exponential tower function defined by $\tower(0) = 1$ and $\tower(m) = 2^{\tower(m-1)}$. 

\begin{theorem} \label{thm:ere}
    Let $\varepsilon > 0$ be an arbitrary constant.  For ERE matching (even for star-free ERE matching), the following hold:
    \begin{enumerate}
    \renewcommand{\theenumi}{(\alph{enumi})}
    \renewcommand{\labelenumi}{\textup{\theenumi}}
        \item \label{thm:ere-ov} Unless moderate-dimension OVC fails, no algorithm runs in $n^{2-\varepsilon} \tower(\asympo(m/\log{m}))$ time.
        \item \label{thm:ere-clique} Unless the $k$-Clique Hypothesis fails, no algorithm runs in $n^{\omega-\varepsilon} \tower(\asympo(m/\log{m}))$ time.  Moreover, unless the Combinatorial $k$-Clique Hypothesis fails, no combinatorial algorithm runs in $n^{3-\varepsilon} \tower(\asympo(m/\log{m}))$ time.
    \end{enumerate}
\end{theorem}

\cref{thm:ere} significantly strengthens \cref{thm:semiere-rsq} for ERE matching.
It follows from \cref{thm:ere}\ref{thm:ere-clique} that, assuming the $k$-Clique hypotheses, any $n^{\omega-\varepsilon}g(m)$ algorithm or $n^{3-\varepsilon}g(m)$ combinatorial algorithm for ERE matching necessarily requires $g$ to be nonelementary (i.e.,~not bounded by any fixed-height exponential tower).
In this sense, as with semi-ERE matching, the approach of Hopcroft and Ullman was already (conditionally) optimal assuming the Combinatorial $k$-Clique Hypothesis and the improvement by \cite{DBLP:journals/corr/abs-2510-09311} using fast matrix multiplication is optimal in this sense assuming the $k$-Clique Hypothesis.
Together with our result for the semi-ERE case (i.e., Theorem~\ref{thm:semiere-rsq}), this sheds light on the mystery of why Hopcroft and Ullman's approach from 1979 has stood the test of time for more than 45 years.

Similar to the cases of semi-ERE and RSQ, if we allow the running time to be nonelementary in $m$, ERE matching can be solved in linear time in $n$ by converting a given ERE to an equivalent NFA of $\tower(\asympO(m))$ states and running Thompson's algorithm.\footnote{While our \cref{thm:ere}\ref{thm:ere-clique} implies that this method requires nonelementary running time in $m$ assuming the hypotheses, the fact actually holds unconditionally because any algorithm that converts a given ERE to an equivalent NFA requires nonelementary running time~\cite{DBLP:conf/stoc/StockmeyerM73,fuererNichtelementareUntereSchranken1978}.}  Thus, \cref{thm:ere}\ref{thm:ere-clique} gives a tight lower bound for the problem in that it establishes the conditionally optimal exponents of $n$ in the running times of the form $n^c \cdot g(m)$ for ERE matching based on whether $g$ is elementary or not.

Many previous studies claimed to have proposed an $\asympO(n^2 m)$-time ERE or semi-ERE matching algorithm~\cite{hirst1989new,DBLP:conf/mfcs/Yamamoto00,DBLP:conf/mfcs/KupfermanZ02,DBLP:conf/stacs/IlieSY03}, which would be a significant improvement over Hopcroft and Ullman's algorithm.  However, all of these claims were later found to be incorrect; see \cite{rosu2005tech,DBLP:conf/fossacs/Rosu07} for more details.
In fact, such an upper bound contradicts our lower bound from \cref{thm:semiere-rsq} unless the $k$-Clique Hypothesis fails or $\omega = 2$.
Moreover, Rosu claimed to have proposed a combinatorial ERE matching algorithm that runs in $\asympO(n^2 \cdot (\log{n} + m) \cdot 2^m)$ time~\cite{DBLP:conf/fossacs/Rosu07}.  However, this bound contradicts our $n^{3-\varepsilon} \tower(\asympo(m/\log{m}))$ lower bound from \cref{thm:ere} and therefore one of the two must be incorrect (or the Combinatorial $k$-Clique Hypothesis is false).
We resolve this inconsistency by identifying an issue with their algorithm.  Specifically, we show that the data structure assumed by the algorithm does not exist in general by constructing a counterexample pair of a string $w$ and an ERE $r$ for which no such data structure exists.  Notably, in constructing this counterexample, we use the encoding technique used in the proof of \cref{thm:ere}.

\begin{remark} \label{rem:chain}
    By combining reductions from previous work, we can obtain $\asympO((n+m)^{\omega-\varepsilon})$ and combinatorial $\asympO((n+m)^{3-\varepsilon})$ lower bounds for semi-ERE matching (and hence also for ERE matching) under the $k$-Clique hypotheses, which are weaker than our bounds in \cref{thm:semiere-rsq}\ref{thm:semiere-rsq-clique} (and \cref{thm:ere}\ref{thm:ere-clique}).  In particular, they are too weak to establish optimality of existing upper bounds or a discrepancy with them in the way discussed above. In more detail, Petersen gave a combinatorial reduction from Greibach's \emph{hardest context-free language} $L_0$~\cite{DBLP:journals/siamcomp/Greibach73} to semi-ERE matching to prove that semi-ERE matching is LOGCFL-complete~\cite{DBLP:conf/stacs/Petersen02}.  Here, the term ``hardest'' refers to the fact that every context-free language (CFL) may be obtained as an inverse homomorphic image of $L_0$ (up to the empty string)~\cite{DBLP:journals/siamcomp/Greibach73}.  
Combining these with the combinatorial reduction by Abboud, Backurs and Vassilevska Williams~\cite{DBLP:journals/siamcomp/AbboudBW18} from $3k$-Clique to the membership problem for a fixed CFL $L$, we have the chain
    \[
        \text{$3k$-Clique} \le L \le L_0 \le \text{semi-ERE matching},
    \]
    and this gives the lower bounds mentioned above. 
    Notably, our reductions underlying \cref{thm:semiere-rsq,thm:rsq-logcfl,thm:ere} are obtained in a unified manner through refining and generalizing this chain by introducing certain novel intermediate problems.
    See \cref{sec:intro-teov} for a detailed discussion.
\end{remark}

\subsubsection{Hardness of Matching with Backreference}
\label{sec:intro-res-rewb}

We refer to regular expressions extended with backreference as \emph{rewbs}.
We only provide an informal semantics of rewbs (see, e.g.,~\cite{freydenberger2019deterministic} for a formal treatment).
Backreference consists of two constructs: a \emph{capturing group} $(r)_i$ to assign a label $i$ to a string that $r$ matches and a \emph{reference} $\bs i$ to denote an expression that matches only the string labeled $i$. 
For example, the language of $(a^*)_1 \bs 1 \bs 1$ is $\{ a^{3n} \mid n \ge 0 \}$ and that of $(a^*)_1 (b^*)_2 \bs 2 \bs 1$ is $\{ a^n b^{2m} a^n \mid n, m \ge 0 \}$.  A rewb with at most $k$ capturing groups is called a \emph{k-rewb}.
As noted earlier, $k$-rewb matching can be solved in $\asympO(n^{2k+1} m)$ time for every $k \ge 1$.

\begin{theorem} \label{thm:ov-rewb}
    Let $k \ge 1$ be a constant.  Unless moderate-dimension $2k$-OVC fails, no $k$-rewb matching algorithm runs in $n^{2k-\varepsilon}\poly(m)$ time (with log-dimension $2k$-OVC, $n^{2k-\varepsilon} 2^{\asympo(\sqrt{m})}$ time) for any constant $\varepsilon > 0$.
\end{theorem}

Recently and independently of our work, Kumabe and Uezato~\cite[Theorem 5]{DBLP:conf/icalp/KumabeU26} showed a conditional lower bound ruling out $\asympO(n^{2k-\varepsilon})$-time $k$-rewb matching algorithms assuming (moderate-dimension) $2k$-OVC for every constant $k \ge 1$.
While their result may appear stronger than \cref{thm:ov-rewb}, their result actually relies on using a non-standard form of rewbs that allows self-referential updates to capturing groups, which is unsupported in almost all practical regular expression engines.\footnote{As far as we are aware, only the .NET regular expression engine supports their rewbs~\cite{dotnetbackref}.}
By contrast, our \cref{thm:ov-rewb} relies only on basic forms of rewbs that are commonly supported by such engines, including those in the standard libraries of Java, Python, and JavaScript, making our result more practically relevant. Furthermore, for $k = 1$, the proof of \cref{thm:ov-rewb} can be rewritten using only \emph{straight-line 1-rewbs}, which are rewbs of the form $e_0 (e)_1 e_1 \bs 1 \cdots e_{l} \bs 1 e_{l+1}$ where $e_0, e, e_1, \dots, e_l, e_{l+1}$ are pure regular expressions (i.e., regular expressions without extensions). The rewbs are practically relevant in that many rewbs that occur in practice are known to be of this form even with $l=1$~\cite{DBLP:conf/mfcs/NogamiT25}.
\begin{theorem} \label{thm:ov-rewb-sl1}
    Unless moderate-dimension OVC fails, no straight-line 1-rewb matching algorithm runs in $n^{2-\varepsilon}\poly(m)$ time (with log-dimension OVC, $n^{2-\varepsilon} 2^{\asympo(\sqrt{m})}$ time) for any constant $\varepsilon > 0$.
\end{theorem}

A recent study by Uezato~\cite{DBLP:conf/cpm/Uezato26} gives an $\asympO(n^2 m^2)$-time algorithm for matching straight-line 1-rewbs, improving the dependence on $n$ of the $\asympO(n^3 m)$ bound obtained by setting $\capn = 1$ in the above $\asympO(n^{2\capn+1} m)$ upper bound for $\capn$-rewb matching.
Therefore, \cref{thm:ov-rewb-sl1} gives a tight lower bound for matching rewbs of this form in that the exponent $2$ of $n$ cannot be shaved by any constant while preserving the factor from $m$ to be polynomial, assuming OVC.

\begin{table}[t]
    \centering
    \caption{Summary of upper and lower bounds for regular expression matching with one or more of the extensions considered in this paper.  SERE denotes semi-EREs and ERE\,+\,RSQ (resp.~SERE\,+\,RSQ) denotes regular expression matching extended with both complement (resp.~intersection) and squaring.}
    \label{tab:sum}
    {
    \renewcommand{\arraystretch}{1.1}
    %
    \begin{tabular}{lll}
        \toprule
        Extension(s)                     & Upper bound            & Notes \\
        \midrule
        ERE\,+\,RSQ                      & $\asympO(n^\omega m)$  & \cite{DBLP:journals/corr/abs-2510-09311} \\
                                         & $\asympO(n^3 m)$       & \cite{hopcroft1979book}, combinatorial \\
                                         & $n \tower(\asympO(m))$ & see \cref{sec:intro-res-ere} \\
        SERE\,+\,RSQ                     & $n 2^{\asympO(m)}$     & see \cref{sec:intro-res-semiere-rsq} \\
                                         & LOGCFL                 & \cite{DBLP:conf/stacs/Petersen02} \\
        $k$-Rewb                         & $\asympO(n^{2k+1} m)$  & \cite{schmid2016characterising,DBLP:journals/jalc/Schmid24,DBLP:conf/mfcs/NogamiT25} \\
        Straight-line 1-rewb             & $\asympO(n^2 m^2)$     & \cite{DBLP:conf/cpm/Uezato26} \\
        \bottomrule
    \end{tabular}
    
    \bigskip
    \begin{tabular}{llll}
        \toprule
        Extension                                               & Lower bound                                         & Assumption       & Notes \\
        \midrule
        SERE\,/\,Star-free RSQ                                  & $n^{\omega-\varepsilon} 2^{\asympo(m)}$             & $k$-Clique       & Thm~\ref{thm:semiere-rsq} \\
                                                                & $n^{3-\varepsilon} 2^{\asympo(m)}$                  & Comb.~$k$-Clique & Thm~\ref{thm:semiere-rsq}, combinatorial \\
                                                                & $n^{2-\varepsilon} 2^{\asympo(m)}$                  & Log-dim OVC      & Thm~\ref{thm:semiere-rsq} \\
                                                                & $n^{2-\varepsilon} \poly(m)$                        & Mod-dim OVC      & Thm~\ref{thm:semiere-rsq} \\
                                                                & LOGCFL-hard                                         & \textemdash      & \cite{DBLP:conf/stacs/Petersen02}\,\&\,Thm~\ref{thm:rsq-logcfl} \\
        Star-free ERE                                           & $n^{\omega-\varepsilon} \tower(\asympo(m/\log{m}))$ & $k$-Clique       & Thm~\ref{thm:ere} \\
                                                                & $n^{3-\varepsilon} \tower(\asympo(m/\log{m}))$      & Comb.~$k$-Clique & Thm~\ref{thm:ere}, combinatorial \\
                                                                & $n^{2-\varepsilon} \tower(\asympo(m/\log{m}))$      & Mod-dim OVC      & Thm~\ref{thm:ere} \\
                                                                & P-hard                                              & \textemdash      & \cite{conf/dcagrs/Petersen00} \\
        $k$-Rewb ($k = \asympO(1)$)                         & $\asympO(n^{2k-\varepsilon})$                       & Mod-dim $2k$-OVC & \cite{DBLP:conf/icalp/KumabeU26}; see Sec~\ref{sec:intro-res-rewb} \\
                                                                & $n^{2k-\varepsilon} 2^{\asympo(\sqrt{m})}$          & Log-dim $2k$-OVC & Thm~\ref{thm:ov-rewb} \\
                                                                & $n^{2k-\varepsilon} \poly(m)$                       & Mod-dim $2k$-OVC & Thm~\ref{thm:ov-rewb} \\
        Straight-line 1-rewb                                    & $n^{2-\varepsilon} 2^{\asympo(\sqrt{m})}$           & Log-dim OVC      & Thm~\ref{thm:ov-rewb-sl1} \\
                                                                & $n^{2-\varepsilon} \poly(m)$                        & Mod-dim OVC      & Thm~\ref{thm:ov-rewb-sl1} \\
        \bottomrule
    \end{tabular}
    }
\end{table}

\cref{tab:sum} summarizes the upper and lower bounds discussed above for regular expression matching extended with one or more of intersection, squaring, complement and backreference.

\subsection{Technical Overview}
\label{sec:intro-teov}

All lower bound results shown in this paper are of the form $n^{c-\varepsilon} g(m)$ and therefore do not rule out algorithms whose running time grows asymptotically faster than $g(m)$ (e.g.,~$\asympO(n^{c-1} 2^m)$ time when $g(m)$ is either $\poly(m)$ or $2^{\asympo(m)}$).  Nevertheless, as discussed in \cref{sec:intro-res}, such bounds suffice to rule out algorithms that are efficient with respect to both $n$ and $m$.   
Moreover, restricting the time complexity of semi-ERE, RSQ and ERE matching with respect to $m$ is even \emph{necessary} for deriving a meaningful lower bound on the complexity with respect to $n$ and $m$ because we could improve the complexity to be linear in $n$ if we were to allow the complexity to be exponential or nonelementary in $m$ as discussed in \cref{sec:intro-res-semiere-rsq,sec:intro-res-ere}.  
    Indeed, investigating this form of multivariate lower bounds has often proven fruitful and has also been pursued in prior work.
    For example, Abboud, Vassilevska Williams and Wang~\cite{DBLP:conf/soda/AbboudWW16} proved that under log-dimension OVC, for any $\varepsilon > 0$, no $n^{2-\varepsilon}2^{\asympo(\tw)}$-time algorithm decides if the diameter of a given graph of treewidth $\tw$ is no more than $2$.
    Additionally, Abboud, Bringmann, Hermelin and Shabtay~\cite{DBLP:journals/talg/AbboudBHS22} proved that under SETH, for any $\varepsilon > 0$, no $T^{1-\varepsilon}2^{\asympo(n)}$-time algorithm solves Subset Sum on $n$ numbers and target $T$.

A key idea in our reductions is to exploit the restricted dependence on $m$ to allow the lengths of the regular expressions produced by the reductions to depend on the sizes of the instances to be reduced.  To illustrate this, we consider a reduction from the Orthogonal Vectors (OV) problem to regular expression matching with extensions, where OV asks whether given sets $A$ and $B$ each containing $N$ Boolean vectors of dimension $d$ have a pair of orthogonal vectors $a \in A$ and $b \in B$.
Given such sets $A$ and $B$, we encode their vectors into a string $w$ and construct a regular expression $r$ that nondeterministically guesses two vectors $a \in A$ and $b \in B$ from $w$ and checks their orthogonality.  In proving a lower bound under log-dimension OVC, we may assume $d = \asympTheta(\log{N})$. 
Accordingly, for the $n^{2-\varepsilon}2^{\asympo(m)}$ and $n^{2-\varepsilon}2^{\asympo(\sqrt{m})}$ lower bounds, we may allow $r$ to have length $\asympO(d)$ and $\asympO(d^2)$, respectively. 
As we shall show, such $r$ can be constructed by exploiting the restricted-dependence-on-$m$ property and the respective regular expression extensions.

For ERE matching, we take this idea further to show the much stronger $n^{2-\varepsilon} \tower(\asympo(m/\log{m}))$ lower bound.  Namely, we construct a reduction from OV to ERE matching that produces extremely short EREs by using the complement extension.  The idea is as follows.  
To decide if the two guessed vectors $a \in A$ and $b\in B$ are orthogonal, we need to check that their values $a[i]$ and $b[i]$ multiply to zero at each position $i \in \{1, \dots, d\}$.  Because the complement operator is available, this can be done by constructing an ERE that checks if ``there is a position $i$ such that $a[i] = b[i] = 1$'' and taking the complement of the ERE.
To this end, in constructing $w$, we encode each value $v[i]$ of a vector $v$ together with the binary representation $\mybin(i)$ of each position $i$ of fixed length $\asympO(\log{d})$.  Then, we construct an ERE that guesses positions $i$ and $j$ of $a$ and $b$, respectively, and checks if $\mybin(i) = \mybin(j)$ and $a[i] = b[j] = 1$.  Further using complement, we can also check if $\mybin(i) = \mybin(j)$ by constructing an ERE that guesses positions $i'$ and $j'$ of $\mybin(i)$ and $\mybin(j)$ respectively, and checks if $\mybin(i') = \mybin(j')$ and $\mybin(i)[i'] \neq \mybin(j)[j']$ and taking its complement.  After repeating this $\asympO(\log^*{d})$ times, we only need to construct an ERE that compares strings of constant length, and this can be done by using intersection.  Consequently, we obtain an ERE of length $\asympO((\log^*{d})\log(\log^*{d}))$ for checking the orthogonality (where the additional $\log(\log^*{d})$ factor arises for technical reasons), and the $n^{2-\varepsilon} \tower(\asympo(m/\log{m}))$ lower bound follows. As mentioned in \cref{sec:intro-res-ere}, we also use this encoding technique to construct our counterexample to the algorithm of \cite{DBLP:conf/fossacs/Rosu07}.

We derive our results for semi-ERE, RSQ and ERE matching (\cref{thm:semiere-rsq,thm:rsq-logcfl,thm:ere}) within a unified framework building on the reduction chain discussed in \cref{rem:chain}.
Recall that the reduction chain yields only the $\asympO((n+m)^{\omega-\varepsilon})$ and combinatorial $\asympO((n+m)^{3-\varepsilon})$ lower bounds for semi-ERE matching under the $k$-Clique hypotheses, too weak for establishing the optimality of the $\asympO(n^\omega m)$ and combinatorial $\asympO(n^3 m)$ upper bounds for the problem.
Our key observation is that the bottleneck lies in the middle reduction $L \le L_0$ and the formulation of $L_0$ which are too general and do not preserve enough structure of $L$. To address this, we ``refine'' $L_0$ into a novel problem that we call \emph{Dyck Selection (DS)} by introducing parameters that retain useful structural information of $L$.
Additionally, we naturally extend DS to \emph{Orthogonal Dyck Selection (ODS)} so as to also accommodate a reduction from OV in addition to those from $k$-Clique and CFL.    
We then go step further and introduce \emph{Generalized Dyck Selection (GDS)}, which subsumes both DS and ODS as special cases.

\begin{figure}
    \centerline{
\begin{tikzpicture}[
    >={Stealth[length=2mm]},
    every node/.style={font=\normalsize},
    baseline=(current bounding box.north),
    node distance=1.5cm and 3.25cm
]

\node (ov) {OV};
\node[base right=of ov] (clique) {$3k$-Clique};
\node[base right=of clique] (cfl) {CFL};
\coordinate (clique-cfl) at ($(clique)!0.5!(cfl)$);
\node[below=of clique-cfl, yshift=0.45cm] (ds) {\textbf{Dyck Selection}};
\coordinate (clique-ds) at ($(clique |- ds)$);
\node[below=of clique-ds, yshift=0.45cm] (ods) {\textbf{Orthogonal Dyck Selection}};
\coordinate (ov-ods) at ($(ov |- ods)$);
\coordinate (gds-top) at ($(ov-ods)!0.5!(ods)$);
\node[below=of gds-top, yshift=0.45cm] (gds) {\textbf{Generalized Dyck Selection}};
\coordinate (ov-gds) at ($(ov |- gds)$);
\node[below=of ov-gds, yshift=0.45cm] (semiere) {Semi-ERE Matching};
\node (ere) at ($(semiere -| clique)$) {ERE Matching};
\node (rsq) at ($(semiere -| cfl)$) {RSQ Matching};

\draw[->] (ds) -- node[midway, right, xshift=2mm] {\footnotesize \cref{lem:ds-ods}} (ods);
\draw[->] (ov) -- node[midway, left, xshift=-2mm] {\footnotesize \cref{lem:ov-ods}} (ods);
\draw[->] (clique) -- node[midway, left, xshift=-1mm] {\footnotesize \cref{lem:clique-ds}} (ds);
\draw[->] (cfl) -- node[midway, right, xshift=2mm] {\footnotesize \cref{lem:cfl-ds}} (ds);
\draw[->] (ods) -- node[midway, left, xshift=-4mm] {} (gds);
\draw[->] (gds) -- node[midway, left, xshift=-4mm] {\footnotesize \cref{lem:ods-semiere}} (semiere);
\draw[->] (gds) -- node[midway, right, xshift=4mm] {\footnotesize \cref{lem:ods-ere}} (ere);
\draw[->] (ods) -- node[midway, right, xshift=4mm] {\footnotesize \cref{lem:ods-rsq}} (rsq);
\end{tikzpicture}
}
    \caption{Overview of the reductions in \cref{sec:ods}.}
    \label{fig:red}
\end{figure}

As indicated in \cref{fig:red}, DS, ODS and GDS serve as useful intermediate problems.  Indeed, as we shall show in \cref{sec:ods}, by modifying Petersen's LOGCFL-hardness reduction for semi-ERE matching~\cite{DBLP:conf/stacs/Petersen02} in a certain way so as to keep the length $m$ of the regular expression in the output instance small, we obtain reductions from GDS to semi-ERE and ERE matching and a reduction from ODS to RSQ matching that yield the desired conditional lower bounds under OVC and the $k$-Clique hypotheses, as well as their LOGCFL-hardness.  As such, these problems, DS, ODS and GDS, formally capture the essence of why the matching problems for semi-ERE, RSQ and ERE are hard, and we foresee them to be also useful for investigating the complexity of other problems, particularly ones that concern formal language theory or stringology.

The rest of the paper is organized as follows.
\cref{sec:prelim} defines preliminary notions.  
\cref{sec:ods} introduces the new intermediate problems DS, ODS and GDS and presents our results on semi-ERE, RSQ and ERE matching by proving the reductions shown in \cref{fig:red}.
\cref{sec:rosu} reviews the algorithm of \cite{DBLP:conf/fossacs/Rosu07} and presents our counterexample.
\cref{sec:ov-rewb} presents our results on rewb matching.

\section{Preliminaries}
\label{sec:prelim}

For integers $i,j$, we write $\intv{i}{j}$ for the set $\{i,i+1,\dots,j\}$.
All logarithms are base 2.  For integer $x \ge 1$, we let $\log^{(0)} x = x$, and for $i \ge 1$, let $\log^{(i)} x = \log(\log^{(i-1)}{x})$.  We denote by $\log^* x$ the minimum $i \ge 0$ such that $\log^{(i)}{x} \le 1$.
For a string or a vector $s$ and an integer $i \ge 1$, we write $\rev{s}$ for the reverse of $s$, $|s|$ for the length of $s$ and $s[i]$ for the $i$-th coordinate of $s$.
For an integer $i \ge 0$, we write $s^i$ for $s$ concatenated $i$ times.  We assume that $\rev{(-)}$, $(-)[i]$ and $(-)^i$ have higher precedence than concatenation.  For example, $s_1 \rev{s_2}$ denotes $s_1(\rev{s_2})$, $s_1 s_2[i]$ denotes $s_1 (s_2[i])$ and $s_1 s_2^i$ denotes $s_1 (s_2^i)$ for strings $s_1$ and $s_2$.
We usually omit the string concatenation symbol, but occasionally write it explicitly using $\odot$.
The syntax of \emph{(pure) regular expressions} over an alphabet $\Sigma$ is given by the following grammar:
\[
    r ::= a \mid \empstr \mid \emptyset \mid r_1 r_2 \mid r_1 \regor r_2 \mid r^*
\]
where $a \in \Sigma$ and $\empstr$ denotes the empty string.  
As with strings, we use $\odot$ as the concatenation symbol when necessary.  For a regular expression $r$ and an integer $i \ge 0$, we write $r^i$ for $r$ concatenated $i$ times.  We assume the standard operator precedence: Kleene star $(-)^*$, concatenation $\odot$ and alternation $\regor$ (in decreasing order).  We also assign $(-)^i$ the same precedence as Kleene star, although we do not regard $(-)^i$ as a primitive operator in regular expressions.  For example, $r_1 r_2^i$ denotes $r_1 (r_2^i)$ for regular expressions $r_1$ and $r_2$.
For brevity, we use the \emph{character set notation} $[a_1 a_2 \cdots a_k]$, which is a shorthand for the regular expression $a_1 \regor a_2 \regor \cdots \regor a_k$ where $a_1, \dots, a_k \in \Sigma$.  For any set $S \subseteq \Sigma$, we regard $S$ as the character set consisting of all elements of $S$.
The \emph{language} of a regular expression is defined in the standard way.  A regular expression $r$ \emph{matches} a string $w$ if $w$ is an element of the language of $r$, which is also written as $w \in r$ by a slight abuse of notation.  Two regular expressions are \emph{equivalent} if they define the same language.
LOGCFL is the class of all decision problems reducible to a context-free language under log-space reductions.

Let $k \ge 2$ be an integer.
The $k$-Orthogonal Vectors ($k$-OV) problem is the following: given $k$ sets $A_1, \dots, A_k \subseteq \{0,1\}^d$ containing the same number $n$ of Boolean vectors of dimension $d$, decide if there are $k$ vectors $a_1 \in A_1, \dots, a_k \in A_k$ such that they are orthogonal (i.e., $\sum_{l=1}^{d} \prod_{i=1}^{k} a_{i}[l] = 0$).
A naive algorithm solves $k$-OV in $\asympO(n^k d)$ time.

\begin{conjecture*}[$k$-Orthogonal Vectors Conjecture, see \cite{DBLP:conf/soda/WilliamsY14,DBLP:conf/soda/ChenW19,DBLP:conf/innovations/KaneW19,DBLP:conf/focs/Williams24} for $k = 2$, and {\cite[Hypothesis 24]{DBLP:journals/corr/abs-2204-11681}} for general $k$] \label{hyp:kov}
    For every constant $\varepsilon > 0$, there is a constant $c \ge 1$ such that $k$-OV cannot be solved in $\asympO(n^{k-\varepsilon})$ time on instances with $d = c\log{n}$.
\end{conjecture*}

In particular, $2$-OV and $2$-OVC are simply called OV and OVC, respectively.
As mentioned in the introduction, it is known that SETH implies $k$-OVC~{\cite{DBLP:journals/tcs/Williams05,DBLP:conf/soda/WilliamsY14,DBLP:journals/corr/abs-2204-11681}}, and therefore bounds obtained by assuming $k$-OVC also hold when assuming SETH.
Moreover, for $k=2$, OVC is supported by evidence beyond their implication from SETH.  See, for example, \cite{DBLP:conf/soda/ChenW19,DBLP:conf/innovations/KaneW19}.

In this paper, we assume log-dimension $k$-OVC in order to derive more refined lower bounds with respect to $m$.  If one is content with polynomial-in-$m$ bounds, assuming the following moderate-dimension version of $k$-OVC is enough.

\begin{conjecture*}[Moderate-dimension $k$-OVC, see~\cite{williams2018some,DBLP:journals/talg/GaoIKW19}] \label{hyp:mdkov}
    For every constant $\varepsilon > 0$, $k$-OV cannot be solved in $n^{k-\varepsilon} \poly(d)$ time.
\end{conjecture*}

Let $k \ge 3$ be an integer.
The $k$-Clique problem asks whether a given undirected graph $G$ has a $k$-clique (i.e.,~the complete graph on $k$ vertices) as a subgraph.  
For a fixed $k$, a naive algorithm solves $k$-Clique in $\asympO(n^k)$ time.  Abboud, Fischer and Shechter~\cite{DBLP:conf/latin/AbboudFS24} have proposed a polylogarithmically faster combinatorial algorithm for the problem.
\begin{hypothesis*}[Combinatorial $k$-Clique Hypothesis, see \cite{DBLP:journals/siamcomp/AbboudBW18,DBLP:conf/latin/AbboudFS24}] \label{hyp:cch}
    For any integer $k \ge 3$ and any constant $\varepsilon > 0$, no combinatorial algorithm can solve $k$-Clique on a graph of $n$ vertices in $\asympO(n^{k - \varepsilon})$ time.
\end{hypothesis*}
Using fast matrix multiplication, Ne{\v{s}}et{\v{r}}il and Poljak showed a faster algorithm that runs in $\asympO(n^{\omega \lceil k/3 \rceil})$ time~\cite{nevsetvril1985complexity}.  Here, $2 \le \omega < 2.3714$~\cite{DBLP:conf/soda/AlmanDWXXZ25} denotes the exponent of square matrix multiplication.

\begin{hypothesis*}[$k$-Clique Hypothesis, see \cite{DBLP:journals/siamcomp/AbboudBW18,williams2018some}] \label{hyp:ch}
    For any integer $k \ge 3$ and any constant $\varepsilon > 0$, no algorithm can solve $k$-Clique on a graph of $n$ vertices in $\asympO(n^{\omega k / 3 - \varepsilon})$ time.  
\end{hypothesis*}

\section{OV, Clique and CFL to Semi-ERE, RSQ and ERE Matching}
\label{sec:ods}

In this section, we prove \cref{thm:semiere-rsq,thm:rsq-logcfl,thm:ere}.
To this end, we introduce the novel intermediate problems Dyck Selection (DS), Orthogonal Dyck Selection (ODS) and Generalized Dyck Selection (GDS) in \cref{sec:ds-ods}, and present reductions to and from them in the subsequent sections.

\subsection{GDS, DS and ODS}
\label{sec:ds-ods}

Following \cite{DBLP:journals/ita/Kemp96,DBLP:journals/siamcomp/Liebehenschel03}, we first present an abstract framework for a generalization of the \emph{Dyck language}, a basic formal language consisting of strings of properly nested parentheses (see, e.g., \cite{hopcroft1979book}).
Given a set $\Delta$ of opening parentheses with the closing parentheses set $\mybar{\Delta} = \{\mybar{x} \mid x \in \Delta\}$ and a binary relation $R$ on $\Delta$, the \emph{generalized Dyck language with respect to $\Delta$, $\mybar{\Delta}$ and $R$} is defined as the set $\{w \mid w \dyckequiv_R \empstr \}$, where $\dyckequiv_R$ is the smallest congruence on $(\Delta \cup \mybar{\Delta})^*$ generated by the relation $\{ (x \mybar{y}, \empstr) \mid x, y \in \Delta,  x R y \}$.

Next, we instantiate this framework as follows.
For a constant $\dt \ge 1$, let $\Gamma_\dt = \{ \ldyck{1}, \dots, \ldyck{\dt} \}$ and $\mybar{\Gamma}_\dt = \{ \rdyck{i} \mid \ldyck{i} \in \Gamma_\dt \}$.
For notation convenience, we often write $\Gamma$ and $\mybar{\Gamma}$ for $\Gamma_\dt$ and $\mybar{\Gamma}_\dt$, respectively.
For an integer $\blu \ge 1$, let $\vecset_{\blu}(\Gamma)$ be the set of $\blu$-dimensional vectors over $\Gamma$ and $\vecset_{\blu}(\mybar{\Gamma})$ the set of those over $\mybar{\Gamma}$.  For $x \in \vecset_{\blu}(\Gamma)$, we write $\mybar{x}$ for the $\beta$-dimensional vector such that $\mybar{x}[i] = \mylongbar{x[i]}$ for each $i \in [1,\blu]$.
Given a binary relation $R$ on $\Gamma$,
we define $\dyckset_R$ as the generalized Dyck language with respect to $\vecset_\blu(\Gamma)$, $\vecset_\blu(\mybar{\Gamma})$ and $R$ lifted to $\vecset_\blu(\Gamma)$, i.e., $x R y \iff \forall i \in \intv{1}{\blu}\colon x[i]\,R\,y[i]$.
In particular, if $p = 1$, $\blu = 1$ and $R = \{(\ldyck{1}, \ldyck{1})\}$, then $\dyckset_R$ coincides with the Dyck language $D_1$.
    
A string in $\dyckset_1$ can naturally be viewed as a rooted tree as follows.
Every $w \in \dyckset_1$ has a unique decomposition $w = \ldyck{1} w_1 \rdyck{1} \cdots \ldyck{1} w_\ell \rdyck{1}$ for some $\ell \ge 0$ and $w_1, \dots, w_\ell \in \dyckset_1$ (see \cref{lem:decomp-aux} in \cref{app:pf-ods-semiere} for the proof).
Using this decomposition, we associate $w$ with a rooted tree $\dycktree(w)$, inductively defined as the tree whose root has $\dycktree(w_1), \dots, \dycktree(w_\ell)$ as its subtrees in this order, with $\dycktree(\empstr)$ consisting of a single leaf.
We define the \emph{depth} of $w\in\dyckset_1$ to be the maximum depth among the vertices of $\dycktree(w)$, i.e., the height of $\dycktree(w)$.
We also define the \emph{degree} of $w \in \dyckset_1$ to be the maximum degree among the vertices of $\dycktree(w)$, where the degree of a vertex in a rooted tree is defined as the number of its children.
For example, $()(())(()())$ has depth $2$ and degree $3$, where the opening and closing parentheses denote $\ldyck{1}$ and $\rdyck{1}$, respectively.
For integers $\dep, \dg \ge 0$, let $\dyckset_1^{(\dep)}$ be the set of strings in $\dyckset_1$ of depth at most $\dep$, and $\dyckset_1^{\depdg{\dep}{\dg}}$ those of depth at most $\delta$ and degree at most $\dg$.

Let $\bq_\blu\colon (\vecset_\blu(\Gamma) \cup \vecset_\blu(\mybar{\Gamma}))^* \to (\Gamma_1 \cup \mybar{\Gamma}_1)^*$ be the homomorphism defined by $\bq_\blu(\vecset_\blu(\Gamma)) = \{\ldyck{1}\}$ and $\bq_\blu(\vecset_\blu(\mybar{\Gamma})) = \{ \rdyck{1} \}$.
We are ready to define our general problem GDS and two of its special cases, DS and ODS.

\begin{definition}
    \label{def:gds-ds-ods}
    Let $X \subseteq \dyckset_1$, which we call the \emph{restriction set}.
    \begin{itemize}
        \item \textbf{Generalized Dyck Selection (GDS).} Let $R$ be a binary relation on $\Gamma$. Given $n$ sets $S_1, \dots, S_n$ of strings over $\vecset_\blu(\Gamma) \cup \vecset_\blu(\mybar{\Gamma})$, decide if there exist $s_1, \dots, s_n$ with $s_i \in S_i$ for every $i$ such that $s_1 \cdots s_n \in \bq_\blu^{-1}(X) \cap \dyckset_R$.\footnote{GDS is defined using $D_R$, the generalized Dyck language with the vector alphabets $\Delta = \vecset_\blu(\Gamma)$, $\mybar{\Delta} = \vecset_\blu(\mybar{\Gamma})$ and a component-wise binary relation on these vectors, rather than the one with arbitrary $\Delta$, $\mybar{\Delta}$ and a binary relation between them.  We do this so that GDS itself can be incorporated into our reduction framework along with its special cases DS and ODS (see \cref{fig:red} in \cref{sec:intro-teov}).}
        \item \textbf{Dyck Selection (DS).} The special case of Generalized Dyck Selection where $p=2$ and $R = \mathord{\releq} = \{ (\ldyck{1}, \ldyck{1}), (\ldyck{2}, \ldyck{2}) \}$.  We denote $\dyckset_{\releq}$ by $\dyckset_2$.
        \item \textbf{Orthogonal Dyck Selection (ODS).} The special case of Generalized Dyck Selection where $p=2$ and $R = \mathord{\relorth} = \{ (\ldyck{1}, \ldyck{1}), (\ldyck{1}, \ldyck{2}), (\ldyck{2}, \ldyck{1}) \}$.
We denote $\dyckset_{\relorth}$ by $\disjset_2$.
            For notation convenience, we write $\ldisj{0}, \ldisj{1}, \rdisj{0}, \rdisj{1}$ instead of $\ldyck{1}, \ldyck{2}, \rdyck{1}, \rdyck{2}$ when discussing this problem.
    \end{itemize}
\end{definition}

Note that setting $X = \dyckset_1$ imposes no restriction in either problem (see \cref{cor:indyck1} in \cref{app:pf-prelim}).
For both problems, we define the input size $N$ by $N = \sum_{i=1}^{n} \sum_{s \in S_i} \max\{\blu \cdot |s|,1\}$.  See \cref{sec:intro-teov} for the motivation for considering these problems.

In what follows, we show the reductions indicated in \cref{fig:red} of \cref{sec:intro-teov}.
By definition, GDS is at least as hard as ODS and DS.
We next show that ODS is at least as hard as DS.

\begin{lemma} \label{lem:ds-ods}
    Let $X \subseteq \dyckset_1$.
    Suppose Orthogonal Dyck Selection with restriction set $X$ can be solved in time $T(N, \blu)$.  Then Dyck Selection with restriction set $X$ can be solved in time $\asympO(N) + T(2N, 2\blu)$ using constant additional space.
\end{lemma}
\begin{proof}
    Let $h\colon (\vecset_\blu(\Gamma) \cup \vecset_\blu(\mybar{\Gamma}))^* \to (\vecset_{2\blu}(\Gamma) \cup \vecset_{2\blu}(\mybar{\Gamma}))^*$ be the unique homomorphism obtained by applying the following substitutions component-wise to each vector symbol:
    \[
        \ldyck{1} \mapsto \ldisj{0}\ldisj{1}, \quad \ldyck{2} \mapsto \ldisj{1}\ldisj{0}, \quad \rdyck{1} \mapsto \rdisj{1}\rdisj{0}, \quad \rdyck{2} \mapsto \rdisj{0}\rdisj{1}.
    \]
    For example, when $\blu = 2$, $h((\ldyck{1}, \ldyck{2}) (\rdyck{1}, \rdyck{2})) = (0,1,1,0) (\mybar{1},\mybar{0},\mybar{0},\mybar{1})$.

    Given a DS instance $S_1, \dots, S_n \subseteq (\vecset_\blu(\Gamma) \cup \vecset_\blu(\mybar{\Gamma}))^*$ of size $N$, convert it into an ODS instance $h(S_1), \dots, h(S_n) \subseteq (\vecset_{2\blu}(\Gamma) \cup \vecset_{2\blu}(\mybar{\Gamma}))^*$ of size $2N$, where $h(S_i) = \{h(s) \mid s \in S_i \}$.  This can be done in $\asympO(N)$ time and constant space.
    Let $w \in (\vecset_\blu(\Gamma) \cup \vecset_\blu(\mybar{\Gamma}))^*$.
    Observe that $\bq_\blu(w) = \bq_{2\blu}(h(w))$ and hence $\pi_\blu(w) \in X \iff \pi_{2\blu}(h(w)) \in X$.
    We claim that $w \in D_2 \iff h(w) \in J_2$ by induction on the length of $w$ (see \cref{app:pf-ds-ods} for details).
    Therefore, DS can be solved in time $\asympO(N) + T(2N, 2\blu)$ by solving the ODS instance.
\end{proof}

\subsection{OV, Clique and CFL to DS and ODS}
\label{sec:ov-clique-ods}

Next, we show reductions from OV, $3k$-Clique and every context-free language to our problems.

\begin{lemma} \label{lem:ov-ods}
    Suppose Orthogonal Dyck Selection with some restriction set containing $\{ \ldyck{1} \rdyck{1} \}$ can be solved in time $T(N, \blu)$.  Then OV can be solved in time $O(nd) + T(2nd, d)$.
\end{lemma}
\begin{proof}
    We construct a reduction from OV to ODS.  Given sets $A$ and $B$ of $d$-dimensional vectors with $|A| = |B| = n$, we encode them into sets $S_1 = A \subseteq \vecset_d(\Gamma)$ and $S_2 = \{ \mybar{y} \mid y \in B \} \subseteq \vecset_d(\mybar{\Gamma})$.  The ODS instance on $S_1$ and $S_2$ has size $2nd$ and can be constructed in $\asympO(nd)$ time.
    Because $\bq_d(x \mybar{y}) = \ldyck{1} \rdyck{1}$ for any $x \in S_1$ and $\mybar{y} \in S_2$, it suffices to take $\{\ldyck{1}\rdyck{1}\}$ as the restriction set for this instance.
    It is easy to see that $x \in A$ and $y \in B$ are orthogonal if and only if $x \mybar{y} \in \disjset_2$.
    Therefore, OV can be solved in time $O(nd) + T(2nd, d)$ by solving the ODS instance.
\end{proof}

\begin{corollary} \label{cor:ov-ods}
    If Orthogonal Dyck Selection with some restriction set containing $\{ \ldyck{1} \rdyck{1} \}$ can be solved in $N^{2-\varepsilon} \poly(\blu)$ time for some constant $\varepsilon > 0$, then moderate-dimension OVC is false.  If it can be solved in $N^{2-\varepsilon} 2^{\asympo(\blu)}$ time for some constant $\varepsilon > 0$, then log-dimension OVC is false.
\end{corollary}
\begin{proof}
    Fix a constant $\varepsilon > 0$ and suppose that ODS with some restriction set containing $\{ \ldyck{1} \rdyck{1} \}$ can be solved in $N^{2-\varepsilon} g(\blu)$ time.
    If $g(\blu) = \poly(\blu)$, then by \cref{lem:ov-ods}, OV can be solved in $n^{2-\varepsilon} \poly(d)$ time, refuting moderate-dimension OVC.
    If $g(\blu) = 2^{\asympo(\blu)}$ and $d = c \log{n}$ for some constant $c \ge 1$, then OV can be solved in $n^{2-\varepsilon+\asympo(1)}$ time, refuting log-dimension OVC.
\end{proof}

\begin{lemma} \label{lem:clique-ds}
    Let $k \ge 1$.  Suppose Dyck Selection with some restriction set containing $\dyckset_1^{\depdg{\asympO(k^2)}{4}}$ can be solved in time $T(N, \blu)$.  Then $3k$-Clique can be solved in time 
    \[
        \asympO(k^3 n^{k+1}) + T(\asympO(k^3 n^{k+1} \log{n}), \asympO(\log{n})). 
    \]
\end{lemma}
\begin{proof}
    We show a combinatorial reduction from $3k$-Clique to DS by adopting the approach of Abboud, Backurs and Vassilevska Williams~\cite{DBLP:journals/siamcomp/AbboudBW18} who showed a combinatorial reduction from $3k$-Clique to the recognition problem for context-free languages.  As in their reduction, we list all $k$-cliques in a given graph and check if there are three disjoint $k$-cliques that are fully connected with each other.  

    Given a graph $G$ of $n$ vertices, we fix $\blu = \asympO(\log{n})$ and identify its vertices with distinct vectors in $\vecset_{\blu}(\Gamma)$.
Let $\mathcal{C}_k$ be the set of all $k$-cliques of $G$.
Fix an arbitrary ordering of each $k$-clique $t$ (e.g.,~the increasing order of vertex indices), and denote the $i$-th vertex in $t$ by $t[i]$.
We identify each $t \in \mathcal{C}_k$ with the length-$k$ string $t[1] \cdots t[k]$ over $\vecset_{\blu}(\Gamma)$.
Let $\vgad(t) = t[1]^k \cdots t[k]^k$.
Define the following sets of strings over $\vecset_{\blu}(\Gamma) \cup \vecset_{\blu}(\mybar{\Gamma})$:
\begin{gather*}
    S_{1} = \{ \vgad(t) \vgad(t) \mid t \in \mathcal{C}_k \}, \ 
    S_{2} = \{ t \mid t \in \mathcal{C}_k \}, \ 
    S_{3} = \{ \rev{\mybar{t}}\, \vgad(t) \mid t \in \mathcal{C}_k \}, \ 
    S_{4} = \{ \rev{\mybar{t}} \mid t \in \mathcal{C}_k \}, \\
    N_{i} = \{ \rev{\mybar{t}} \mybar{u} t \mid u \in \nb(t[i]), t \in \mathcal{C}_k \} \text{ for $i \in \intv{1}{k}$, where $N(t[i])$ is the neighborhood of $t[i]$,}
\end{gather*}
and form a sequence $\mathcal{S}$ of these sets
    \[
        \mathcal{S} = S_1, S_2, \underbrace{\mathcal{N}, \dots, \mathcal{N}}_{\text{$k$ copies}}, S_3, S_2, \underbrace{\mathcal{N}, \dots, \mathcal{N}}_{\text{$k$ copies}}, \underbrace{\mathcal{N}, \dots, \mathcal{N}}_{\text{$k$ copies}}, S_4
    \]
    where $\mathcal{N} = N_1, \dots, N_k$.
    The DS instance on $\mathcal{S}$ has size $\asympO(k^3 n^{k+1} \log{n})$ and can be constructed in $\asympO(k^3 n^{k+1})$ time.
    Observe that it suffices to take $\dyckset_1^{\depdg{\asympO(k^2)}{4}}$ as the restriction set for this instance.
    We claim that $G$ has a $3k$-clique if and only if this is a yes-instance (see \cref{app:pf-clique-ds} for details).
    Therefore, $3k$-Clique can be solved in the claimed time by solving the DS instance.
\end{proof}

\begin{corollary} \label{cor:clique-ds}
    If, for some constant $\varepsilon > 0$, Dyck Selection with some restriction set containing $\dyckset_1^{\depdg{\dep}{4}}$ can be solved in $N^{\omega-\varepsilon} 2^{\asympo(\blu)}$ time for every constant $\dep > 0$, then the $k$-Clique Hypothesis is false.  
    If, for some constant $\varepsilon > 0$, it can be solved combinatorially in $N^{3-\varepsilon} 2^{\asympo(\blu)}$ time for every constant $\dep > 0$, then the Combinatorial $k$-Clique Hypothesis is false.
\end{corollary}
\begin{proof}
    Let $c \in \{ \omega, 3\}$.  Fix a constant $\varepsilon > 0$ and suppose that DS with some restriction set containing $D_1^{\depdg{\dep}{4}}$ can be solved in $N^{c-\varepsilon} 2^{\asympo(\blu)}$ time for every constant $\dep > 0$.  
Let $k$ be a sufficiently large constant such that $c < (k+1) \varepsilon$.
By \cref{lem:clique-ds}, we can solve $3k$-Clique in time $n^{(k+1)(c-\varepsilon) + \asympo(1)}$, which is $\asympO(n^{ck-\varepsilon'})$ for some $\varepsilon' > 0$.  Because the reduction is combinatorial, a combinatorial algorithm remains combinatorial after the reduction, completing the proof. 
\end{proof}

\begin{lemma} \label{lem:cfl-ds}
    Every context-free language is (simultaneous) linear-time and constant-space reducible to Dyck Selection.
\end{lemma}
\begin{proof}
    The proof is almost identical to that of \cite{DBLP:journals/siamcomp/Greibach73} (see \cref{app:pf-cfl-ds} for details).
\end{proof}

\begin{corollary} \label{cor:cfl-ds}
    Dyck Selection and Orthogonal Dyck Selection with restriction set $\dyckset_1$ are LOGCFL-hard, and Generalized Dyck Selection with restriction set $\dyckset_1$ is in LOGCFL for any binary relation $R$.  In particular, Dyck Selection and Orthogonal Dyck Selection are LOGCFL-complete.
\end{corollary}
\begin{proof}
    The LOGCFL-hardness of DS follows from \cref{lem:cfl-ds}.  By \cref{lem:ds-ods}, ODS is also LOGCFL-hard.
    See \cref{app:pf-cfl-ds} for a proof that GDS is in LOGCFL.
\end{proof}

\subsection{GDS to Semi-ERE Matching and ODS to RSQ Matching}
\label{sec:ods-semiere-rsq}

Next, we show reductions from Generalized Dyck Selection to semi-ERE matching and from Orthogonal Dyck Selection to RSQ matching.
Semi-ERE~\cite{hopcroft1979book} extends regular expressions with the \emph{intersection} construct $r_1 \regand r_2$.  The language of a semi-ERE is defined by letting the language of $r_1 \regand r_2$ be the intersection of the languages of $r_1$ and $r_2$.
For example, $(aa)^* \cap (aaa)^*$ is equivalent to $(aaaaaa)^*$, whose language is $\{ a^{6n} \mid n \ge 0 \}$.
We assign intersection $\regand$ the same precedence as alternation $\regor$.

\begin{lemma} \label{lem:ods-semiere}
    Suppose semi-ERE matching can be solved in time $T(n,m)$.  Then Generalized Dyck Selection with restriction set $D_1^{(\dep)}$ can be solved in time $\asympO(N + \blu \dep) + T(\asympO(N), \asympO(\blu \dep))$.
\end{lemma}
\begin{proof}
    We construct a reduction from GDS to semi-ERE matching.  Our construction closely follows that of Petersen~\cite[Lemma 4]{DBLP:conf/stacs/Petersen02}.  Given a GDS instance $S_1, \dots, S_n \subseteq (\vecset_\blu(\Gamma) \cup \vecset_\blu(\mybar{\Gamma}))^*$, we encode it as 
    \[
        w = \bigodot_{i=1}^{n} \left( \left( \bigodot_{s \in S_i} \hash \vectostr(s) \right) \hash \dollar \right),
    \]
where $\vectostr\colon (\vecset_\blu(\Gamma) \cup \vecset_\blu(\mybar{\Gamma}))^* \to (\Gamma^\blu \cup \mybar{\Gamma}^\blu)^*$ is the homomorphism defined by $\vectostr(x) = x[1] \cdots x[\blu]$ and $\vectostr(\mybar{x}) = \rev{\mylongbar{\vectostr(x)}}$ for $x\in \vecset_\blu(\Gamma)$.
    Both the length of $w$ and the time required for this construction are $\asympO(N)$.  We then define a semi-ERE $r$ over the alphabet $\Sigma = \Gamma \cup \mybar{\Gamma} \cup \{ \hash, \dollar \}$ as follows:
    \begin{align*}
        r &= (\Sigma\setminus\{\dollar\})^* \hash r_\dep \hash (\Sigma\setminus\{\dollar\})^* \dollar, \\
        r_i &= r_0 (\rzeroi{\blu}\dia{r_{i-1}} r_0)^* \text{ for } i \in \intv{1}{\dep}, \\
        r_{0} &= \hash (\Sigma \setminus \{\dollar\})^* \dollar (\Sigma \setminus \{\dollar\})^* \hash + \empstr, \\
        \rzeroi{j}\dia{r'} &= \Gamma \, \rzeroi{j-1}\dia{r'} \, \mybar{\Gamma} \cap r_R \text{ for } j \in \intv{1}{\blu}, \\
        r_{R} & = a_1 \Sigma^* \mybar{b}_1 \regor \cdots \regor a_k \Sigma^* \mybar{b}_k \text{ where } R = \{(a_1, b_1), \dots, (a_k,b_k)\}, \\
        \rzeroi{0}\dia{r'} &= r'.
    \end{align*}
    Both the length of $r$ and the time required for this construction are $\asympO(\blu \dep)$.
    We claim that the given GDS instance $S_1, \dots, S_n$ is a yes-instance if and only if $r$ matches $w$ (see \cref{app:pf-ods-semiere} for details).  Therefore, the given GDS instance can be solved in the claimed time by solving the semi-ERE matching instance.
\end{proof}

We next consider RSQ matching.
RSQ~\cite{DBLP:conf/focs/MeyerS72} extends regular expressions with the \emph{squaring} construct $\regsq{r}$, which denotes an expression equivalent to $rr$. 
We assign squaring $\regsq{(-)}$ the same precedence as Kleene star.

\begin{lemma} \label{lem:ods-rsq}
    Suppose RSQ matching can be solved in time $T(n,m)$.  Then Orthogonal Dyck Selection with restriction set $\dyckset_1^{(\dep)}$ can be solved in time $\asympO(N + \blu \dep) + T(\asympO(N), \asympO(\blu \dep))$.
\end{lemma}

\begin{proof}
    We construct a reduction similar to that in \cref{lem:ods-semiere}, but with squaring instead of intersection.
    Given an ODS instance $S_1, \dots, S_n \subseteq (\vecset_\blu(\Gamma) \cup \vecset_\blu(\mybar{\Gamma}))^*$, we encode it as 
    \[
        w = \bigodot_{i=1}^{n} \left( \left( \bigodot_{s \in S_i} \hash \myenc(\vectostr(s)) \right) \hash \dollar \right),
    \]
    where $\vectostr\colon (\vecset_\blu(\Gamma) \cup \vecset_\blu(\mybar{\Gamma}))^* \to (\Gamma^\blu \cup \mybar{\Gamma}^\blu)^*$ is the homomorphism defined in the proof of \cref{lem:ods-semiere} and $\myenc\colon (\Gamma_2 \cup \bar{\Gamma}_2)^* \to \{1, \cent_1, \cent_2, \mybar{\cent}_1, \mybar{\cent}_2\}^*$ is the homomorphism defined by $\myenc(0) = \cent_1 1 \cent_2$, $\myenc(1) = \cent_1 11 \cent_2$, $\myenc(\mybar{0}) = \mybar{\cent}_2 1 \mybar{\cent}_1$ and $\myenc(\mybar{1}) = \mybar{\cent}_2 11 \mybar{\cent}_1$.
    Both the length of $w$ and the time required for this construction are $\asympO(N)$.  We then define an RSQ $r$ over the alphabet $\Sigma = \{ 1, \cent_1, \cent_2, \mybar{\cent}_1, \mybar{\cent}_2, \hash, \dollar \}$ as follows:
    \begin{align*}
        r &= (\Sigma\setminus\{\dollar\})^* \hash r_\dep \hash (\Sigma\setminus\{\dollar\})^* \dollar, \\
        r_i &= r_0 (\rzeroi{\blu}\dia{r_{i-1}} r_0)^* \text{ for } i \in \intv{1}{\dep}, \\
        r_{0} &= \hash (\Sigma \setminus \{\dollar\})^* \dollar (\Sigma \setminus \{\dollar\})^* \hash + \empstr, \\
        \rzeroi{j}\dia{r'} &= \cent_1 \regsq{(1\cent_2 \,\rzeroi{j-1}\dia{r'}\, \mybar{\cent}_2 1 + 1 + \empstr)} \mybar{\cent}_1 \text{ for } j \in \intv{1}{\blu}, \\
        \rzeroi{0}\dia{r'} &= r'.
    \end{align*}
    (Only the definition of $\rzeroi{j}\dia{r'}$ differs from that in \cref{lem:ods-semiere}.)
    Both the length of $r$ and the time required for this construction are $\asympO(\blu \dep)$.
    We claim that the given ODS instance $S_1, \dots, S_n$ is a yes-instance if and only if $r$ matches $w$ (see \cref{app:pf-ods-rsq} for details).  Therefore, the given ODS instance can be solved in the claimed time by solving the RSQ matching instance.
\end{proof}

\begin{corollary} \label{cor:ods-rsq-starfree}
    Suppose star-free RSQ matching can be solved in time $T(n,m)$.  Then Orthogonal Dyck Selection with restriction set $\dyckset_1^{(\dep)}$ can be solved in time 
    \[
        \asympO(N + (\blu+\log{N}) \dep) + T(\asympO(N), \asympO((\blu+\log{N}) \dep))
    \]
    using $\asympO(\log{\blu}+\log{\dep}+\log{N})$ additional space.
\end{corollary}
\begin{proof}
    We avoid using the Kleene star in the construction of $r$ by adopting a technique used in \cite[Lemma 2]{DBLP:conf/stacs/Petersen02}.  
    Let $e^*$ be a starred subexpression of $r$. Instead of $e^*$, we use
    \[
        \regsq{(\cdots (\regsq{(\regsq{(e\regor\empstr)})}) \cdots )}
    \]
    where squares are nested $\lceil \log{|w|} \rceil$ times.  Then, $r$ matches $w$ if and only if it matches $w$ after this change, because the matching is determined by which subexpressions of $r$ match which substrings of $w$ and the lengths of the latter are at most $|w|$.  Both the length of $r$ and the time required for the construction increase to $\asympO((\blu + \log{N})\dep)$.
    The additional $\asympO(\log{\blu}+\log{\dep}+\log{N})$ space is required for counters needed for the construction of $r$ and the above technique.
\end{proof}

\cref{thm:rsq-logcfl} follows from \cref{lem:cfl-ds,cor:ods-rsq-starfree} by setting $\dep = N$.
We next prove \cref{thm:semiere-rsq}.

\begin{proof}[Proof of \cref{thm:semiere-rsq}]
    Let $c \ge 2$ be a constant and $g$ a function to be chosen later.
    Suppose that semi-ERE matching on a string of length $n$ and a semi-ERE of length $m$ can be solved in $n^{c-\varepsilon} g(m)$ time for some $\varepsilon > 0$.
    By choosing $\varepsilon$ sufficiently small, we may assume that $c - \varepsilon \ge 1$.
    
    Set $c = 2$.  
    By \cref{lem:ods-semiere}, we can solve ODS with restriction set $\dyckset_1^{(1)}$ in $N^{2-\varepsilon} g(\asympO(\blu))$ time.
    Because $\{ \ldyck{1} \rdyck{1} \} \subseteq \dyckset_1^{(1)}$, \cref{cor:ov-ods} refutes moderate-dimension OVC if $g(m) = \poly(m)$, and it refutes log-dimension OVC if $g(m) = 2^{\asympo(m)}$.  These prove part (a) of the theorem.

    For (b), set $g(m) = 2^{\asympo(m)}$.  
    By \cref{lem:ds-ods,lem:ods-semiere}, we can solve DS with restriction set $\dyckset_1^{(\dep)}$ for any constant $\dep > 0$ in $N^{c-\varepsilon} 2^{\asympo(\blu)}$ time.
    Because $\dyckset_1^{\depdg{\dep}{4}} \subseteq \dyckset_1^{(\dep)}$, \cref{cor:clique-ds} with $c = \omega$ and $c = 3$ refutes the $k$-Clique hypotheses.
    The proof for star-free RSQ matching is analogous, using \cref{cor:ods-rsq-starfree}.
\end{proof}

\begin{remark} \label{rem:semiere-unbdd}
    If one is content with the weaker $n^{c-\varepsilon} 2^{\asympo(\sqrt{m})}$ lower bound, we can show the hardness of semi-ERE matching with a single use of unbounded intersection.  
    Indeed, it is easy to see that $\rzeroi{\blu}\dia{r'}$ in the proof of \cref{lem:ods-semiere} is equivalent to $\rzeroi{\blu}'\dia{r'} = (\bigcap_{j=1}^{\blu} \Gamma^{j-1} r_R \mybar{\Gamma}^{j-1}) \cap \Gamma^\blu r' \mybar{\Gamma}^\blu$, which uses unbounded intersection exactly once and no other intersection.
    When $\rzeroi{\blu}'\dia{r'}$ is used instead of $\rzeroi{\blu}\dia{r'}$, the length of $r$ and its construction time become $\asympO(\blu^2)$.  Therefore, the resulting bound is $n^{c-\varepsilon} 2^{\asympo(\sqrt{m})}$.  
    Moreover, the $n^{2-\varepsilon} \poly(m)$ lower bound under moderate-dimension OVC still holds after this modification.
\end{remark}

\subsection{GDS to ERE Matching}
\label{sec:ods-ere}

Next, we show a reduction from Generalized Dyck Selection to ERE matching.
ERE~\cite{hopcroft1979book} extends regular expressions with the \emph{complement} construct $\regnot r$.  The language of $\regnot r$ is the complement of the language of $r$, that is, the set of strings over the alphabet that are not matched by $r$.  For example, $\regnot ((aa)^*)$ is equivalent to $a(aa)^*$ over the alphabet $\{ a \}$.  
We assign complement $\regnot$ the same precedence as Kleene star.
We use the intersection $r_1 \regand r_2$ as a shorthand for $\regnot (\regnot r_1 \regor \regnot r_2)$.

\begin{lemma} \label{lem:ods-ere}
    Suppose star-free ERE matching can be solved in time $T(n,m)$.  Then Generalized Dyck Selection with restriction set $\dyckset_1^{\depdg{\dep}{\dg}}$ can be solved in time
    \[
        N \blu^{1+\asympo(1)} + \asympO(\dg^\dep (\log^*{\blu}) \log(\log^*{\blu})) + T(N \blu^{1+\asympo(1)}, \asympO( \dg^\dep (\log^*{\blu})  \log(\log^*{\blu}))).
    \]
\end{lemma}

\begin{proof}
    We construct a reduction from GDS to star-free ERE matching.
    For integer $x \ge 1$, we define $\ld(x) = \lfloor \log(x) + 1 \rfloor$.  Observe that any integer $1 \le i \le x$ has the binary representation of length $\ld(x)$, which we denote by $\mybin_{\ld(x)}(i)$.  Let $\ld^{(i)}(x) = \ld(\ld^{(i-1)}(x))$ for $i \ge 1$ and $\ld^{(0)}(x) = x$.
    Let $\Sigma = \Gamma \cup \mybar{\Gamma} \cup \{ \cent_1, \cent_2, \percent, \mybar{\cent}_1, \mybar{\cent}_2, \mybar{\percent}, \hash, \dollar \}$.  
    Let $S_1, \dots, S_n \subseteq (\vecset_\blu(\Gamma) \cup \vecset_\blu(\mybar{\Gamma}))^*$ be a given GDS instance.
    If $\blu \le 2$, we may increase $\blu$ to at least $3$ by repeating each character three times, without changing the answer of the instance (this follows by a straightforward induction using \cref{lem:decomp-deg} from \cref{app:pf-ods-ere}).
    We encode the instance as
    \[
        w = \bigodot_{i=1}^{n} \left( \left( \bigodot_{s \in S_i} \hash \myenc_h(\vectostr(s)) \right) \hash \dollar \right),
    \]
    where $\vectostr\colon (\vecset_\blu(\Gamma) \cup \vecset_\blu(\mybar{\Gamma}))^* \to (\Gamma^\blu \cup \mybar{\Gamma}^\blu)^*$ is the homomorphism defined in the proof of \cref{lem:ods-semiere}, $h \ge 1$ is the smallest integer such that $\ld^{(h)}(\blu) \le 2$, and $\myenc_h \colon (\Gamma^\blu \cup \mybar{\Gamma}^\blu)^* \to (\Sigma \setminus \{\hash,\dollar\})^*$ is the homomorphism induced by the map $\myenc_h \colon \Gamma^\blu \cup \mybar{\Gamma}^\blu \to (\Sigma \setminus \{\hash,\dollar\})^*$ defined as
    \begin{align*}
        \myenc_{i}(v) &= \left( \bigodot_{j=1}^{|v|} \bindelim{\cent}{1,i}\,\myenc_{i-1}(\mybin_{\ld(|v|)}(j))\,\bindelim{\cent}{2,i}\,v[j] \right) \bindelim{\percent}{i} \text{\quad for } i \in \intv{1}{h} \text{ and } v \in \Gamma^{\ld^{(h-i)}(\blu)}, \\
                      & \qquad \qquad \text{where } \bindelim{\cent}{1,i} = \cent_1\,\mybin_{\ld(h)}(i),\ \bindelim{\cent}{2,i} = \cent_2\,\mybin_{\ld(h)}(i) \text{ and } \bindelim{\percent}{i} = \percent\,\mybin_{\ld(h)}(i), \\
        \myenc_{0}(v) &= v \text{\quad for } v \in \Gamma^{\ld^{(h)}(\blu)}, \\
        \myenc_i(\mybar{v}) &= \rev{\mylongbar{\myenc_i(v)}} \text{\quad for } i \in \intv{0}{h} \text{ and } v \in \Gamma^{\ld^{(h-i)}(\blu)}.
    \end{align*}
    Note that $h = \asympO(\log^*{\blu})$.  Indeed, we prove by induction that $\tower(i-1) + 2 \le \ld^{(h-i)}(\blu)$ for all $i \in [1,h]$.     The base case $i=1$ holds because $\ld^{(h-1)}(\blu) \ge 3$.
    Assume that the inequality holds for some $i \ge 1$.  Then,
    \[
        \tower(i-1) + 2 \le \ld^{(h-i)}(\blu) = \ld(\ld^{(h-(i+1))}(\blu)) \le \log(\ld^{(h-(i+1))}(\blu)) + 1
    \]
    and hence $\ld^{(h-(i+1))}(\blu) \ge 2^{\tower(i-1)+1} \ge \tower(i)+2$. 
    In particular, $\tower(h - 1) \le \blu$ holds and therefore
    $h = \asympO(\log^*{\blu})$.

    We compute the length of $w$.
    By definition, the length of $\myenc_i(v)$ depends only on the length of $v$.
    Hence, for simplicity, we only present the case $v = 1^\blu$. Then we have
    \begin{align*}
        |\myenc_h(1^\blu)| &\le \blu \cdot |\myenc_{h-1}(1^{\ld(\blu)})| + 4(1+\ld(h)) \blu \\
                           &\le \blu \ld(\blu) \cdot |\myenc_{h-2}(1^{\ld^{(2)}(\blu)})| + 4(1+\ld(h)) (\blu + \blu \ld(\blu)) \\
                           &\le \blu \ld(\blu) \ld^{(2)}(\blu) \cdot |\myenc_{h-3}(1^{\ld^{(3)}(\blu)})| + 4(1+\ld(h)) (\blu + \blu \ld(\blu) + \blu \ld(\blu) \ld^{(2)}(\blu)) \\
                           &\quad \vdots \\
                           &\le \left(\prod_{i=0}^{h-1} \ld^{(i)}(\blu)\right)\cdot |\myenc_{0}(1^{\ld^{(h)}(\blu)})| + 4(1+\ld(h)) \sum_{j=0}^{h-1} \prod_{i=0}^{j} \ld^{(i)}(\blu) \\
                           &\le   9 h \ld(h) \prod_{i=0}^{h} \ld^{(i)}(\blu) = \asympO(\blu (\log^2{\blu}) (\log^*{\blu}) \log(\log^*{\blu})) = \blu^{1+\asympo(1)}.
    \end{align*}
    Here, we use that $\prod_{i = 2}^{h} \ld^{(i)}(\blu) = \asympO(\log{\blu})$~\cite[Exercise 1.3(2)]{DBLP:books/vollmer}.  Therefore, $|w| = N \blu^{1+\asympo(1)}$.   

    Next, we analyze the time required for the construction of $w$.  
    For $v \in \Gamma^\blu$, we construct $\myenc_h(v)$ as follows.
    We compute $\ld^{(i)}(\blu)$, $i \in \intv{1}{h}$, and $h$ by repeatedly counting the number of digits in binary until it is at most $2$.
    This can be done in time
    \[
        \asympO\left(\sum_{i=0}^h \ld^{(i)}(\blu)\right) \le \asympO(\blu) + \asympO\left(\prod_{i=2}^{h} \ld^{(i)}(\blu)\right) = \asympO(\blu) + \asympO(\log{\blu}) = \asympO(\blu).
    \]
    Then, we generate $\myenc_h(v)$ from left to right using a counter $c = (c_h, \dots, c_1)$, with each $c_i$ ranging from $1$ to $\ld^{(h-i)}(\blu)$, and with $c$ incremented in lexicographic order.
    Overall, $\blu^{1+\asympo(1)}$ time suffices to construct $\myenc_h(v)$.
    Therefore, $w$ can be constructed in $N \blu^{1+\asympo(1)}$ time.
   
    Next, we define a star-free ERE $r$ over $\Sigma$ as follows, where we use $\Sigma^*$ as shorthand for $\regnot \emptyset$ and write $\cap$ vertically for readability:
    \begin{align*}
        r &= \rfree{\dollar}\,\hash r_\dep \hash\,\rfree{\dollar}\,\dollar, \\
        r_i &= r_0 (\rzeroi{\blu}\dia{r_{i-1}}\,r_0 + \empstr)^\dg \text{ for } i \in \intv{1}{\dep}, \\
        r_{0} &= \hash \,\rfree{\dollar}\, \dollar \,\rfree{\dollar}\, \hash + \empstr, \\
        \rzeroi{\blu}\dia{r'} &= \regnot
            \left(
                \rfree{\bindelim{\percent}{h}}\,\bindelim{\cent}{1,h}
                \left(
                    \begin{gathered}
                        \reqi{\ld^{(1)}(\blu)} \\
                        \regand \\
                        \rfree{\bindelim{\cent}{2,h}}\,\bindelim{\cent}{2,h}\, \regnot r_R\,\bindelimc{\cent}{2,h}\,\rfree{\bindelimc{\cent}{2,h}}
                    \end{gathered}
                \right)
                \bindelimc{\cent}{1,h}\,\rfree{\bindelimc{\percent}{h}}
            \right) \\
                              & \quad \regand \bindelim{\cent}{1,h}\,\rfree{\bindelim{\percent}{h}}\,\bindelim{\percent}{h}\,r'\,\bindelimc{\percent}{h}\,\rfree{\bindelimc{\percent}{h}}\,\bindelimc{\cent}{1,h}, \\
        r_{R} & = a_1 \Sigma^* \mybar{b}_1 \regor \cdots \regor a_k \Sigma^* \mybar{b}_k \text{ where } R = \{(a_1, b_1), \dots, (a_k,b_k)\}, \\
       \reqi{\ld^{(i)}(\blu)} &= \regnot
            \left(
                \rfree{\bindelim{\percent}{h-i}}\,\bindelim{\cent}{1,h-i}
                \left(
                    \begin{gathered}
                        \reqi{\ld^{(i+1)}(\blu)} \\
                        \regand \\
                        \rfree{\bindelim{\cent}{2,h-i}}\,\bindelim{\cent}{2,h-i} \, \regnot \req \, \bindelimc{\cent}{2,h-i}\,\rfree{\bindelimc{\cent}{2,h-i}}
                    \end{gathered}
                \right)
                \bindelimc{\cent}{1,h-i}\,\rfree{\bindelimc{\percent}{h-i}}
            \right) 
            \tag*{for $i \in \intv{1}{h-1}$,}
            \\
                              & \ \text{where } 
                                    \begin{array}{lll}
                                        \bindelim{\cent}{1,i} = \cent_1\,\mybin_{\ld(h)}(i), & \bindelim{\cent}{2,i} = \cent_2\,\mybin_{\ld(h)}(i), & \bindelim{\percent}{i} = \percent\,\mybin_{\ld(h)}(i),\\
                                        \bindelimc{\cent}{1,i} = \rev{\mylongbar{\bindelim{\cent}{1,i}}}, & \bindelimc{\cent}{2,i} = \rev{\mylongbar{\bindelim{\cent}{2,i}}}, & \bindelimc{\percent}{i} = \rev{\mylongbar{\bindelim{\percent}{i}}},
                                    \end{array}
                               \\
       \reqi{2} &= \Sigma \reqi{1} \Sigma \regand \reqi{1}, \quad \reqi{1} = \req = 0 \Sigma^* \mybar{0} + 1 \Sigma^* \mybar{1}, \\
        \rfree{s} &= \regnot ( \Sigma^* s \Sigma^* ) \text{ for } s \in \Sigma^*.
    \end{align*}
    Note that $\reqi{\ld^{(h)}(\blu)}$ is either $\reqi{1}$ or $\reqi{2}$ because $\ld^{(h)}(\blu) \le 2$.
    A simple calculation gives
    \[
        |r| = \asympO(\dg^\dep h \log{h}) = \asympO(\dg^\dep (\log^*{\blu}) \log(\log^*{\blu})),
    \]
    and $r$ can also be constructed in the same time bound using the value of $h$, which has been already computed during the construction of $w$.
    
    We claim that the given GDS instance $S_1, \dots, S_n$ is a yes-instance if and only if $r$ matches $w$ (see \cref{app:pf-ods-ere} for details).  Therefore, the given GDS instance can be solved in time 
    \[
        N \blu^{1+\asympo(1)} + \asympO(\dg^\dep (\log^*{\blu}) \log(\log^*{\blu})) + T(N \blu^{1+\asympo(1)}, \asympO( \dg^\dep (\log^*{\blu}) \log(\log^*{\blu})))
    \]
    by solving the star-free ERE matching instance.
\end{proof}


We are now ready to prove \cref{thm:ere}.

\begin{proof}[Proof of \cref{thm:ere}]
    The proof is similar to that of \cref{thm:semiere-rsq}, using \cref{lem:ods-ere}.
    Let $c \ge 2$ be a constant to be chosen later.
    Suppose that star-free ERE matching on a string of length $n$ and an ERE of length $m$ can be solved in $n^{c-\varepsilon} \tower(\asympo(m/\log{m}))$ time for some $\varepsilon > 0$.
    By choosing $\varepsilon$ sufficiently small, we may assume that $c - \varepsilon \ge 1$.
    
    By \cref{lem:ods-ere}, we can solve ODS with restriction set $\dyckset_1^{\depdg{\dep}{4}}$ for any constant $\dep > 0$ in time 
    \[
        (N\blu^{1+\asympo(1)})^{c-\varepsilon} \tower\left(\asympo\left( 
            \frac
            {(\log^*{\blu}) \log(\log^*{\blu})}
            {\log(\log^*{\blu}) + \log^{(2)}(\log^*{\blu})}
        \right)\right) = n^{c-\varepsilon} \poly(\blu).
    \]
    As in the proof of \cref{thm:semiere-rsq}, if $c = 2$, \cref{cor:ov-ods} refutes moderate-dimension OVC, and if $c \in \{ \omega, 3\}$, \cref{lem:ds-ods,cor:clique-ds} refute the $k$-Clique hypotheses.
\end{proof}

\section{A Counterexample to the ERE Matching Algorithm of \texorpdfstring{\cite{DBLP:conf/fossacs/Rosu07}}{[Ros07]}}\label{sec:rosu}
In this section, we show that
the $\asympO(n^2 \cdot (\log n + m) \cdot 2^m)$-time ERE matching algorithm proposed in \cite[Theorem 1]{DBLP:conf/fossacs/Rosu07} has a structural flaw,
by providing a counterexample to a fundamental assumption of the algorithm.

Their algorithm is as follows.
Let $w = a_1 \cdots a_n$ be a string, where $a_1, \dots, a_n$ are characters, and let $r$ be an ERE.
The basic idea of the algorithm is to calculate
 the $(n+1) \times (n+1)$ Boolean matrix $T$ such that for any $0 \le i < j \le n$,
 \[ T[i][j] = 1 \quad\iff\quad a_{i+1} \cdots a_{j} \in \regnot r'\]
 for each negated subexpression $\regnot r'$ of $r$.
As explained in \cite[Proposition 3]{DBLP:conf/fossacs/Rosu07}, this gives an ERE matching algorithm that runs in $\asympO(n^3 k + n^2m)$ time, where $k$ is the number of complement operators in $r$.
Then, to obtain a more efficient algorithm with respect to $n$,
a data structure called \emph{jumping machine}
is introduced for expressing the above matrix more compactly with respect to $n$.
\begin{definition*}[{\cite[Definition 6]{DBLP:conf/fossacs/Rosu07}}]
A \emph{jumping machine} $\mathcal{P} = (P, p_0, \pi)$ consists of 
set $P$ of states,
an initial state $p_0$, and
a jumping map $\pi \colon \intv{0}{n-1} \times P \to (\intv{1}{n} \times P) \cup \{\bot\}$ with the property that for any $i \in \intv{0}{n-1}$ and any $p \in P$,
if $\pi(i, p) = (j, p')$, then $i < j$.
Given $i \in \intv{0}{n}$ and $p \in P$,
we let $\pi_{p}(i) \defeq \{j \mid \exists p' \in P.\,\exists l \ge 1.\,\pi^{l}(i, p) = (j, p')\}$. Here, $\pi^{l}(i, p) \defeq \pi^{l-1}(\pi(i, p))$ if $\pi(i, p) \neq \bot$, 
and $\pi^{l}(i, p) \defeq \bot$ otherwise for $l \ge 2$,
and $\pi^{1}(i, p) \defeq \pi(i, p)$.
Given a word $w = a_1 \cdots a_n$ and an ERE $r$,
we say that $\mathcal{P} = (P, p_0, \pi)$ is a \emph{jumping machine for $w$ and $r$} if for any $i \in \intv{0}{n-1}$, we have
\[\pi_{p_0}(i) ~=~ \{j \in \intv{i+1}{n} \mid a_{i+1} \cdots a_{j} \in r\}.\]
\end{definition*}
The $\asympO(n^2 \cdot (\log n + m) \cdot 2^m)$-time ERE matching algorithm is obtained from the algorithm above by replacing Boolean matrices with jumping machines with $\asympO(2^m)$ states~\cite[Theorem 1]{DBLP:conf/fossacs/Rosu07}.
At this point, the following is assumed:
\[
    \text{For any string $w$ and ERE $r$, there is a jumping machine with $\asympO(2^m)$ states for $w$ and $r$.} \tag{$\P$} \label{hypo:rosu}
\]
However, we show that the claim \eqref{hypo:rosu} is incorrect for general EREs.%
\footnote{The claim \eqref{hypo:rosu} holds for a pure regular expression
by taking $P$ as the states of a deterministic finite automaton equivalent to the pure regular expression.}
That is, we provide $w$ and $r$ such that any jumping machine for them must have a number of states that is super-exponential in $m$.
We first prove the following lemma, which presents a lower bound on the number of states of any jumping machines for $w$ and $r$.
\begin{lemma}\label{lem:rosu}
    Let $w = a_1 \cdots a_n$ be a string and $r$ be an ERE.
    Let $j$ be a position such that $a_{i+1} \cdots a_j \in r$ for all $i \in \intv{0}{j-1}$.
    Let $C(w, r, j)$ be the number of distinct sets in the collection 
    $\{\xi(i) \mid i \in \intv{0}{j-1}\}$, where
    \[\xi(i) \defeq \{j' \in \intv{j+1}{n} \mid a_{i+1} \cdots a_{j'} \in r\}.\]
    Then, any jumping machine for $w$ and $r$ must have at least $C(w, r, j)$ states.
\end{lemma}
\begin{proof}
    For each $i \in \intv{0}{j-1}$,
    there exist $p_i \in P$ and $l \ge 1$ such that $\pi^{l}(i, p_0) = (j, p_i)$ because $a_{i+1} \cdots a_j \in r$.
    We claim that $\pi_{p_i}(j) = \xi(i)$.
    This holds because 
    both are obtained by intersecting $\pi_{p_0}(i) = \{j' \in \intv{i+1}{j} \mid \exists p' \in P.\,\exists l \ge 1.\,\pi^{l}(i,p_0) = (j',p')\} \cup \pi_{p_i}(j)$ and $\{ j' \in \intv{i+1}{n} \mid a_{i+1} \cdots a_{j'} \in r \}$ with $\intv{j+1}{n}$.
    Therefore, for each $\xi(i)$, we can choose $p \in P$ such that $\pi_{p}(j) = \xi(i)$.  This defines an injection $\xi(i) \mapsto p$ and hence the cardinality of $P$ is at least $C(w,r,j)$.
\end{proof}

Below, we exhibit $w$, $r$ and $j$ such that $C(w, r, j)$ is at least super-exponential in $m$.%
\footnote{In fact, by recursively applying the same technique of encoding word positions in binary, we can obtain $w, r$ and $j$ such that $C(w, r, j)$ is $k$-fold exponential in $m$ for any fixed $k$.  As mentioned in \cref{sec:intro-res-ere,sec:intro-teov}, this encoding technique was also used in the proofs of \cref{lem:ods-ere}.}
Let $q \ge 1$ be an integer.
Let $w$ be the string over $\{0, 1, \percent, \cent, \hash, \dollar\}$ defined as follows:
\begin{align*}
    w &\defeq \hash \myenc(0) \hash \myenc(1) \cdots \hash \myenc(2^{2^{q}}-1) \dollar \rev{\myenc(2^{2^{q}}-1)} \hash \cdots \rev{\myenc(1)} \hash \rev{\myenc(0)} \hash,\\
    \myenc(i) &\defeq \cent\percent \mybin_{q}(0) \percent \mybin_{2^{q}}(i)[1] \; \cent\percent \mybin_{q}(1) \percent \mybin_{2^{q}}(i)[2] \\
              & \qquad \qquad \qquad \qquad \qquad \qquad \cdots \cent \percent \mybin_{q}(2^{q}-1) \percent \mybin_{2^{q}}(i)[2^{q}] \text{\quad for $i \in \intv{0}{2^{2^{q}}-1}$}.
\end{align*}
Here, for any integer $i \in \intv{0}{2^{q}-1}$ (resp.~$\intv{0}{2^{2^{q}}-1}$), we write $\mybin_{q}(i)$ (resp.~$\mybin_{2^{q}}(i))$ for the binary representation of $i$ of length ${q}$ (resp.~$2^{q}$).
Let $j$ be the position of $\dollar$ in $w$ (i.e., $w[j] = \dollar$).
Let $r$ be the ERE defined as follows:
\begin{align*}
    r &~\defeq~ \Sigma^* \dollar ~\regor~ \hash (\req \regand \Sigma^* \dollar \Sigma^* ) \hash,\\
    \req &~\defeq~ \regnot ([01 \percent\cent]^* {\cent\percent} {\reqi{q}\dia{\percent(0\Sigma^* 1 \regor 1\Sigma^* 0)\percent}} {\percent\cent} [01 \percent\cent]^*), \\
    {\reqi{l}\dia{r'}} &~\defeq~ {\Sigma \reqi{l-1}\dia{r'} \Sigma \cap (0 \Sigma^* 0 \regor 1 \Sigma^* 1) \text{\quad for $l \in \intv{1}{q}$},} \\
    {\reqi{0}\dia{r'}} &~\defeq~ {r'}.
\end{align*}
Observe that the length $m$ of $r$ is $\asympO(q)$.
The ERE $r$ satisfies the following properties:
\begin{enumerate}
    \item \label{rosu:1} For any string $s \in \Sigma^*$, the ERE $r$ matches $s\dollar$.
    \item \label{rosu:2} For any substring $s$ of $w$ of the form $\hash\myenc(i)\, s_1 \dollar s_2 \,\rev{\myenc(i')}\hash$ where $i, i' \in \intv{0}{2^{2^{q}}-1}$ are integers and $s_1, s_2 \in [01\percent\cent\hash]^*$ are strings, $r$ matches $s$ if and only if $i = i'$.
\end{enumerate}
Thus, the behavior of any jumping machine for $w$ and $r$ is as shown in \cref{fig:rosu}.
Using this figure, we prove the following theorem, which refutes the claim \eqref{hypo:rosu}.
\begin{figure}
    \centering
\begin{tikzpicture}[
    >=stealth,
    cyan line/.style={draw=cyan!80!black, thick},
    cyan text/.style={text=cyan!80!black},
    orange box/.style={draw=orange, thick, rounded corners=3pt},
    black box/.style={draw=black, thick, rounded corners=3pt},
    every node/.style={font=\normalsize},
    baseline=(current bounding box.north),
    node distance=0.5cm
]



\matrix (word) [matrix of math nodes, column sep=0.01cm, row sep=0.8cm] at (4, -2) {
    w = & \node(e1h){\hash\myenc(0)}; & \node(e2h){\hash\myenc(1)}; & \node(ld){\cdots}; & \node(e3h){\hash\myenc(2^{2^{q}}-1)}; & \node(d){\dollar}; & \node(e4h){\rev{\myenc(2^{2^{q}}-1)}\hash}; & \node(rd){\cdots}; & \node(e5h){\rev{\myenc(1)}\hash}; & \node(e6h){\rev{\myenc(0)}\hash}; \\
};

\node(i1)[below = of e1h.west]{$i_{1}$};
\node(i2)[below = of e2h.west]{$i_{2}$};
\node(i3)[below = of e3h.west]{$i_{2^{2^{q}}}$};
\node(i23) at ($(i2)!0.5!(i3)$) {$\cdots$};
\node(j)[below = of d.east]{$j$};
\node(i4)[below = of e4h.east]{$i'_{2^{2^{q}}}$};
\node(i5)[below = of e5h.east]{$i'_{2}$};
\node(i45) at ($(i4)!0.5!(i5)$) {$\cdots$};
\node(i6)[below = of e6h.east]{$i'_{1}$};


\draw[->] (i1) -- (e1h.west);
\draw[->] (i2) -- (e2h.west);
\draw[->] (i3) -- (e3h.west);
\draw[->] (j) -- (d.east);
\draw[->] (i4) -- (e4h.east);
\draw[->] (i5) -- (e5h.east);
\draw[->] (i6) -- (e6h.east);

\draw (-3.9, -3.35) -- (12.3, -3.35);



\node (m-1-1a)[below = 1.8cm of e1h.west]{$(i_1, p_0)$};
\node (m-1-2a)[minimum width = 4.5em] at (m-1-1a -| j) {$(j, p_1)$};
\node (m-1-3a)[xshift = -3mm] at (m-1-1a -| i6) {$(i'_1, p'_1)$};
\node (m-2-1a) at ($(m-1-1a -| i2) + (0, -0.75cm)$) {$(i_2, p_0)$};
\node (m-2-2a)[minimum width = 4.5em] at (m-2-1a -| j) {$(j, p_2)$};
\node (m-2-3a) at (m-2-1a -| i5) {$(i'_2, p'_2)$};
\node (m-3-1a) at ($(m-2-1a -| i2) + (1cm, -0.75cm)$) {$\ddots$};
\node (m-3-2a)[minimum width = 4.5em] at (m-3-1a -| j) {$\vdots$};
\node (m-3-3a) at ($(m-2-3a -| i5) + (-1cm, -0.75cm)$) {$\myiddots$};
\node (m-4-1a) at ($(m-3-1a -| i3) + (0, -0.8cm)$) {$(i_{2^{2^{q}}}, p_0)$};
\node (m-4-2a)[minimum width = 4.5em] at (m-4-1a -| j) {$(j, p_{2^{2^{q}}})$};
\node (m-4-3a) at (m-4-1a -| i4) {$(i'_{2^{2^{q}}}, p'_{2^{2^{q}}})$};

\draw[->, thick, shorten >=2pt, shorten <=2pt] (m-1-1a) -- (m-1-2a); 
\draw[->, thick, shorten >=2pt, shorten <=2pt] (m-1-2a) -- (m-1-3a);

\draw[->, thick, shorten >=2pt, shorten <=2pt] (m-2-1a) -- (m-2-2a); 
\draw[->, thick, shorten >=2pt, shorten <=2pt] (m-2-2a) -- (m-2-3a);

\draw[->, thick, shorten >=2pt, shorten <=2pt] (m-4-1a) -- (m-4-2a); 
\draw[->, thick, shorten >=2pt, shorten <=2pt] (m-4-2a) -- (m-4-3a);

\node [black box, fit=(m-1-2a) (m-4-2a), inner xsep=-3.5pt, inner ysep=3pt] {};


\end{tikzpicture}
    \caption{Behavior of a jumping machine for $w$ and $r$ constructed earlier.
    The intermediate states $p_1, \dots, p_{2^{2^{q}}}$ at position $j$ are pairwise distinct because so are the accepting positions $i'_1, \dots, i'_{2^{2^{q}}}$.}
    \label{fig:rosu}
\end{figure}

\begin{theorem}\label{thm:rosu}
    For any function $f(m) = 2^{2^{\asympo(m)}}$,
    there are a string $w$ and an ERE $r$ of length $m$ such that
    any jumping machine for $w$ and $r$ must have at least $f(m)$ states.
    Consequently, the claim \eqref{hypo:rosu} is false.
\end{theorem}
\begin{proof}
Let $w$, $r$ and $j$ be as constructed above.
Let $i_1, \dots, i_{2^{2^{q}}}, j, i'_{2^{2^{q}}}, \dots, i'_1$ be positions in $w$ as shown in \cref{fig:rosu}.
By Property \ref{rosu:2} above, for each ${l, l'} \in \intv{1}{2^{2^{q}}}$,
we have that $i_{{l'}}' \in \xi(i_{l})$ if and only if $l = l'$.
Hence, sets $\xi(i_1)$, $\dots$, $\xi(i_{2^{2^{q}}})$ are pairwise distinct and therefore $C(w, r, j) \ge 2^{2^{q}}$.
Since $m = \asympO(q)$,
it follows that $C(w, r, j) = 2^{2^{\asympOmega(m)}}$.
Therefore, $C(w, r, j) \ge f(m)$ holds for sufficiently large $m$.
By \cref{lem:rosu} and Property \ref{rosu:1} above,
any jumping machine for $w$ and $r$ must have at least $f(m)$ states.
\end{proof}

\section{OV to Rewb Matching}
\label{sec:ov-rewb}

In this section, we prove \cref{thm:ov-rewb,thm:ov-rewb-sl1} by establishing the following \cref{lem:kov-krewb,lem:ov-rewb}.
Recall the informal semantics of rewbs described in \cref{sec:intro-res-rewb}.

\begin{lemma} \label{lem:kov-krewb}
    Let $k \ge 1$ be a constant. 
    Suppose $k$-rewb matching can be solved in time $T(n,m)$.  Then $2k$-OV can be solved in time $\asympO(nd^2) + T(nd^2, d^2)$.
\end{lemma}

\begin{proof}
    We construct a reduction from $2k$-OV to $k$-rewb matching as follows.  Given sets of vectors $A_1, \dots, A_{2k}$ where $A_i = \{ a_{i,j} \mid 1 \le j \le n \}$ for $i \in \intv{1}{2k}$, we encode them into a string $w$ in the following way:
    \[
        w = (s \cent)^{d+1} \text{ where } s = \bigodot_{i = 1}^{k} \left( \left( \bigodot_{j=1}^{n} a_{2i-1,j}\hash \right) \dollar \left(\bigodot_{j=1}^{n} \hash \rev{a_{2i,j}} \right) \dollar \right)
    \]    
    Both the length of $w$ and the time required for this construction are $\asympO(nd^2)$.  We then define a $k$-rewb $r$ over the alphabet $\Sigma = \{ 0, 1, \hash, \dollar, \cent \}$ as follows:
    \begin{align*}
        r &= \left(\bigodot_{i=1}^{k} [01\hash]^* (\hash [01\hash]^* \dollar [01\hash]^* \hash)_i [01\hash]^* \dollar \right) \cent \rzeroi{1} \cent \rzeroi{2} \cent \cdots \cent \rzeroi{d} \cent, \\
        \rzeroi{l} &= \rzeroij{l}{1} \regor \rzeroij{l}{2} \regor \cdots \regor \rzeroij{l}{k} \text{ for } l \in \intv{1}{d}, \\
        \rzeroij{l}{i} &= ([01\hash]^*\dollar)^{2(i-1)}[01\hash]^* (0[01]^{l-1}\bs i \regor \bs i [01]^{l-1}0) [01\hash\dollar]^* \text{ for } i \in \intv{1}{k}.
    \end{align*}
    Both the length of $r$ and the time required for this construction are $\asympO(d^2)$.
    We claim that there exist vectors $a_1 \in A_1, \dots, a_{2k} \in A_{2k}$ that are orthogonal if and only if $r$ matches $w$ (see \cref{app:pf-ov-rewb} for details). Therefore, the given $2k$-OV instance can be solved in the claimed time by solving the $k$-rewb matching instance.
\end{proof}

\begin{lemma} \label{lem:ov-rewb}
    Suppose straight-line 1-rewb matching can be solved in time $T(n,m)$.  Then OV can be solved in time $\asympO(nd^2) + T(nd^2, d^2)$.
\end{lemma}

\begin{proof}
    We construct a reduction from OV to straight-line 1-rewb matching as follows.  Given sets of vectors $A = \{ a_1, \dots, a_n \}$ and $B = \{ b_1, \dots, b_n \}$, we encode them into a string $w$ in the following way: 
    \[
        w = (s \cent)^{d+1} \text{ where } s = a_1 \hash a_2 \hash \cdots a_n \hash \dollar \hash \rev{b_1} \hash \rev{b_2} \cdots \hash \rev{b_n}.
    \]
    Both the length of $w$ and the time required for this construction are $O(nd^2)$.
    We then define a straight-line 1-rewb $r$ over the alphabet $\Sigma = \{ 0,1,\hash, \dollar, \cent\}$ as follows:
    \begin{align*}
        r &= [01\hash]^* (\hash \Sigma^* \hash)_1 \rzeroi{1} \bs 1 \rzeroi{2} \bs 1 \cdots \rzeroi{d} \bs 1 [01\hash]^* \cent, \\
        \rzeroi{l} &= [01]^{l-1} 0[01\hash]^* \cent [01\hash]^* \regor [01\hash]^*\cent[01\hash]^*0[01]^{l-1} \text{ for } l \in \intv{1}{d}.
    \end{align*}
    Both the length of $r$ and the time required for this construction are $O(d^2)$.
    We claim that there exist vectors $a_i \in A$ and $b_j \in B$ that are orthogonal if and only if $r$ matches $w$ (see \cref{app:pf-ov-rewb} for details). Therefore, the given OV instance can be solved in the claimed time by solving the straight-line 1-rewb matching instance.
\end{proof}

\cref{thm:ov-rewb,thm:ov-rewb-sl1} follow immediately from the above lemmas.

\bibliographystyle{alpha}
\bibliography{ref}

\appendix
\section{Detailed Proofs}
\label[appendix]{app:proof}

\subsection{Proof preliminaries}
\label[appendix]{app:pf-prelim}

We first introduce a useful function (cf.~\cite{Harrison1978}).
In what follows, $R$ denotes a binary relation on $\Gamma$, $x,y$ denote vectors in $\vecset_\blu(\Gamma)$ and $u,v,w$ denote vector strings in $(\vecset_\blu(\Gamma) \cup \vecset_\blu(\mybar{\Gamma}))^*$.
We also often omit the subscript $\blu$, writing $\vecset(\Gamma)$, $\vecset(\mybar{\Gamma})$ and $\bq$ for $\vecset_\blu(\Gamma)$, $\vecset_\blu(\mybar{\Gamma})$ and $\bq_\blu$, respectively.

\begin{definition} \label{def:mu}
Define a map $\mu_R\colon (\vecset(\Gamma) \cup \vecset(\mybar{\Gamma}))^* \to (\vecset(\Gamma) \cup \vecset(\mybar{\Gamma}))^*$ by
\begin{align*}
    \mu_R(\empstr) &= \empstr, \\
    \mu_R(w x) &= \mu(w) x, \\ 
        \mu_R(w \mybar{y}) &= 
            \begin{cases}
                w' & \text{if } \mu(w) = w' x \text{ for some $w'$ and $x \in \vecset(\Gamma)$ with $x R y$}, \\
                \mu(w) \mybar{y} & \text{otherwise}.
            \end{cases}
\end{align*}
When $R$ is clear from the context, we omit the subscript and write $\mu$.
\end{definition}
We will use the following properties of $\mu$.

\begin{lemma}[{cf.~\cite[Propositions 10.4.1 and 10.4.2]{Harrison1978}}]
    \label{lem:mu-harrison}
    The following hold:
    \begin{enumerate}
        \renewcommand{\labelenumi}{\textup{(\alph{enumi})}}
        \renewcommand{\theenumi}{\textup{(\alph{enumi})}}
        \item $\mu(\mu(u)) = \mu(u)$.
        \item $\mu(uv) = \mu(\mu(u)v) = \mu(\mu(u)\mu(v))$.
        \item \label{lem:mu-harrison-rewind} If $\mu(w \mybar{y}) \in \vecset(\Gamma)^*$, then $\mu(w) = \mu(w \mybar{y}) x$ for some $x \in \vecset(\Gamma)$ with $x R y$.
        \item \label{lem:mu-harrison-prefix} If $u$ is a prefix of a string in $\mu^{-1}(\empstr)$, then $\mu(u) \in \vecset(\Gamma)^*$.
    \end{enumerate}
\end{lemma}
\begin{proof}
    The proofs of (a) and $\mu(uv) = \mu(\mu(u)v)$ are analogous to those in \cite[Proposition 10.4.1(e) and (f)]{Harrison1978} and are omitted.
    We prove $\mu(u \mu(v)) = \mu(uv)$ by induction on $|v|$.  The base case $v = \empstr$ is immediate.
    If $v = v' x$ for some $x \in \vecset(\Gamma)$, then 
    \[
        \mu(u \mu(v)) = \mu(u \mu(v') x) = \mu(u \mu(v')) x = \mu(u v') x =\mu(u v' x) = \mu(uv).
    \]
    Suppose $v = v' \mybar{y}$ for some $y \in \vecset(\Gamma)$.  We distinguish two cases.
    If $\mu(v') = v'' x$ for some $x \in \vecset(\Gamma)$ with $x R y$, then $\mu(v) = v''$. So 
    \[
        \mu(uv') = \mu(u \mu(v')) = \mu(u v'' x) = \mu(u v'') x = \mu(u \mu(v)) x
    \]
    and hence $\mu(uv) = \mu(uv' \mybar{y}) = \mu(u \mu(v))$.
    Otherwise, we have $\mu(v) = \mu(v') \mybar{y}$.  So
    \[
        \mu(u\mu(v)) = \mu(u \mu(v') \mybar{y}) = \mu(\mu(u\mu(v'))\mybar{y}) = \mu(\mu(uv')\mybar{y}) = \mu(uv' \mybar{y}) = \mu(uv),
    \]
    completing the induction.  Therefore, $\mu(\mu(u)\mu(v)) = \mu(\mu(u)v) = \mu(uv)$.
    Part (c) follows directly from the definition of $\mu$. 
    For (d), we prove by induction on $|v|$ that $\mu(uv) = \empstr \Rightarrow \mu(u) \in \vecset(\Gamma)^*$ for all $u$.  The base case $v = \empstr$ is immediate.  If $v = xv'$ for some $x \in \vecset(\Gamma)$, then $\mu(ux) = \mu(u)x \in \vecset(\Gamma)^*$ and hence $\mu(u) \in \vecset(\Gamma)^*$.
    If $v = \mybar{y} v'$ for some $y \in \vecset(\Gamma)$, then $\mu(u\mybar{y}) \in \vecset(\Gamma)^*$.
    By (c), $\mu(u) = \mu(u\mybar{y}) x$ for some $x \in \vecset(\Gamma)$, and hence $\mu(u) \in \vecset(\Gamma)^*$.
\end{proof}

\begin{lemma} \label{lem:mudyck}
    $\mu(u) = \mu(v) \iff u \dyckequiv v$.  In particular, $\dyckset_R = \mu^{-1}(\empstr)$.
\end{lemma} 
\begin{proof}
    We write $u \dyckequiv_\mu v$ if $\mu(u) = \mu(v)$.  Note that $\dyckequiv_\mu$ is a congruence containing $S = \{ (x\mybar{y}, \empstr) \mid x,y \in \vecset(\Gamma), xRy \}$.
    Indeed, $\mu(x\mybar{y}) = \empstr$, and if $\mu(u) = \mu(v)$, then for every $w_1,w_2$, we have $\mu(w_1 u w_2) = \mu(\mu(w_1) \mu(u) \mu(w_2)) = \mu(\mu(w_1) \mu(v) \mu(w_2)) = \mu(w_1 v w_2)$.  Hence, $\mathord{\dyckequiv} \subseteq \mathord{\dyckequiv_\mu}$.
    For the converse, we show $u \dyckequiv \mu(u)$ by induction on $|u|$.  
    The base case $u = \empstr$ is immediate.
    If $u = u' x$ for some $u'$ and $x \in \vecset(\Gamma)$, we have $\mu(u) = \mu(u') x \dyckequiv u' x = u$.  
    If $u = u' \mybar{y}$ for some $y \in \vecset(\Gamma)$, we distinguish two cases.
    If $\mu(u') = u'' x$ for some $x \in \vecset(\Gamma)$ with $x R y$, then $\mu(u) = u'' \dyckequiv u'' x \mybar{y} = \mu(u') \mybar{y} \dyckequiv u' \mybar{y} = u$.
    Otherwise, we have $\mu(u) = \mu(u') \mybar{y} \dyckequiv u' \mybar{y} = u$.  
    Therefore, if $\mu(u) = \mu(v)$, then $u \dyckequiv \mu(u) = \mu(v) \dyckequiv v$, and hence $\mathord{\dyckequiv_\mu} \subseteq \mathord{\dyckequiv}$.  This completes the proof.
\end{proof}

\begin{lemma}[{inversion}] \label{lem:inv}
    Every nonempty $w \in \dyckset_R$ can be decomposed as $w = w_1 x w_2 \mybar{y}$ with $w_1, w_2 \in \dyckset_R$ and $x R y$.
\end{lemma}
\begin{proof}
    We prove the claim that if $\mu(w) = x_1 \cdots x_\ell$ for some $\ell \ge 0$ and $x_1, \dots, x_\ell \in \vecset(\Gamma)$, then $w = w_1 x_1 \cdots w_\ell x_\ell w_{\ell+1}$ for some $w_1, \dots, w_{\ell+1} \in \dyckset_R$, by induction on $|w|$.  The base case $w = \empstr$ is immediate.  
    If $w = w' x$ for some $w'$ and $x \in \vecset(\Gamma)$, we have $\mu(w) = \mu(w') x = x_1 \cdots x_\ell$ and hence obtain $x = x_\ell$ and $\mu(w') = x_1 \cdots x_{\ell-1}$.  By the induction hypothesis, $w'$ can be written as $w_1 x_1 \cdots w_{\ell-1} x_{\ell-1} w_\ell$ for some $w_1, \dots, w_\ell \in \dyckset_R$.  So $w = w' x = w_1 x_1 \cdots w_{\ell-1} x_{\ell-1} w_\ell x_\ell w_{\ell+1}$ with $w_{\ell+1} = \empstr$, and the claim holds.
    If $w = w' \mybar{y}$ for some $y \in \vecset(\Gamma)$, by \cref{lem:mu-harrison}\ref{lem:mu-harrison-rewind}, we have $\mu(w') = \mu(w) x = x_1 \cdots x_\ell x$ for some $x$ with $x R y$.  By the induction hypothesis, $w' = w_1 x_1 \cdots w_\ell x_\ell w_{\ell+1} x w_{\ell+2}$ for some $w_1, \dots, w_{\ell+2} \in \dyckset_R$ and hence $w = w_1 x_1 \cdots w_\ell x_\ell w_{\ell+1} x w_{\ell+2} \mybar{y}$.  Because $\mu(w_{\ell+1} x w_{\ell+2} \mybar{y}) = \empstr$, the claim also holds.  This completes the proof of the claim.

    Because $\mu(w) = \empstr$ and $w$ is nonempty, $w$ is of the form $w = w' \mybar{y}$.  By \cref{lem:mu-harrison}\ref{lem:mu-harrison-rewind} and the claim above, $\mu(w') = x$ for some $x$ with $x R y$ and $w' = w_1 x w_2$ for some $w_1, w_2 \in \dyckset_R$.  Therefore, $w = w_1 x w_2 \mybar{y}$, which completes the proof.
\end{proof}

\begin{corollary} \label{cor:indyck1}
    $\dyckset_R \subseteq \bq^{-1}(\dyckset_1)$.
\end{corollary}
\begin{proof}
    By a straightforward induction on $|w|$ using \cref{lem:inv}.
\end{proof}
\subsection{Proof of \texorpdfstring{\cref{lem:ds-ods}}{Lemma \getrefnumber{lem:ds-ods}}}
\label[appendix]{app:pf-ds-ods}

\begin{proof}[Proof of \cref{lem:ds-ods}]
We prove that $w \in \dyckset_2 \iff h(w) \in \disjset_2$ by induction on $|w|$.  
The base case $w = \empstr$ is immediate, so assume $w$ is nonempty.
Suppose that $w \in \dyckset_2$.
By the inversion lemma (\cref{lem:inv}), $w$ can be written as $w = w_1 x w_2 \mybar{x}$ for some $w_1, w_2 \in \dyckset_2$ and $x \in \vecset_\blu(\Gamma)$.
By the induction hypothesis, $\mu_{\relorth}(h(w)) = \mu_{\relorth}(h(x) h(\mybar{x})) = \empstr$ and hence $h(w) \in \disjset_2$.
Conversely, suppose that $h(w) \in \disjset_2$.  
Again, by the inversion lemma, $h(w)$ can be written as $h(w) = h(w_1) h(x) h(w_2) h(\mybar{y})$ for some $w_1, w_2$ with $h(w_1), h(w_2) \in \disjset_2$ and some $x, y \in \vecset_\blu(\Gamma)$ with $h(x) \relorth \mylongbar{h(\mybar{y})}$.\footnote{Here, applying the bar twice cancels out.}  It follows easily that $x = y$.  By the induction hypothesis, $\mu_{\releq}(w) = \mu_{\releq}(x \mybar{y}) = \empstr$ and hence $w \in \dyckset_2$.
\end{proof}
\subsection{Proof of \texorpdfstring{\cref{lem:clique-ds}}{Lemma \getrefnumber{lem:clique-ds}}}
\label[appendix]{app:pf-clique-ds}

\begin{lemma} \label{lem:mu}
    Let $R$ be a binary relation on $\Gamma$ and $u,v,w,x,y \in (\vecset(\Gamma) \cup \vecset(\mybar{\Gamma}))^*$.
    \begin{enumerate}
        \renewcommand{\labelenumi}{\textup{(\alph{enumi})}}
        \item\label{lem:mu-a} If $\mu(v) = \empstr$, then $\mu(uvw) = \mu(uw)$.
        \item\label{lem:mu-b} If $\mu(ux\mybar{y}v) = \empstr$, then $x R y$ and $\mu(uv) = \empstr$.
    \end{enumerate}
\end{lemma}
\begin{proof}
    Immediate from \cref{lem:mu-harrison}. 
\end{proof}

\begin{proof}[Proof of \cref{lem:clique-ds}]
    We show that $G$ has a $3k$-clique if and only if the Dyck Selection instance on $\mathcal{S}$ is a yes-instance.
    Suppose $G$ has a $3k$-clique, or equivalently, three disjoint $k$-cliques $t_1, t_2, t_3$ that are fully connected with each other.  Let $v_{i,j} = t_i[j]$ for $i \in \intv{1}{3}$ and $j \in \intv{1}{k}$.  We can form
    \[
    \begin{aligned}
        s &= \vgad(t_1) \vgad(t_1) 
        t_2 
        (\rev{\mybar{t}_2} \mybar{v}_{1,k} t_2)^k \cdots (\rev{\mybar{t}_2} \mybar{v}_{1,1} t_2)^k
        \rev{\mybar{t}_2} \\
        &\qquad\vgad(t_2)
        t_3
        (\rev{\mybar{t}_3} \mybar{v}_{2,k} t_3)^k \cdots (\rev{\mybar{t}_3} \mybar{v}_{2,1} t_3)^k
        (\rev{\mybar{t}_3} \mybar{v}_{1,k} t_3)^k \cdots (\rev{\mybar{t}_3} \mybar{v}_{1,1} t_3)^k
        \rev{\mybar{t}_3}
    \end{aligned}
    \]
    by choosing one string from each set of $\mathcal{S}$. By \cref{lem:mu}(a), 
    \[
        \mu(s) = \mu\Big(\vgad(t_1)\vgad(t_1) (\mybar{v}_{1,k})^k \cdots (\mybar{v}_{1,1})^k \vgad(t_2) (\mybar{v}_{2,k})^k \cdots (\mybar{v}_{2,1})^k (\mybar{v}_{1,k})^k \cdots (\mybar{v}_{1,1})^k\Big) = \empstr
    \]
    and therefore $s \in \dyckset_2$.
    Conversely, suppose that some $s \in \dyckset_2$ is obtained by choosing one string from each set in $\mathcal{S}$.  Write
    \[
    \begin{aligned}
        s ={}&\vgad(t_1) \vgad(t_1) \\
             &t_2 \\
             &(\rev{\mybar{t}_{3,1,1}} \mybar{u}_{3,1,1} t_{3,1,1} \cdots \rev{\mybar{t}_{3,1,k}} \mybar{u}_{3,1,k} t_{3,1,k})
             \cdots
             (\rev{\mybar{t}_{3,k,1}} \mybar{u}_{3,k,1} t_{3,k,1} \cdots \rev{\mybar{t}_{3,k,k}} \mybar{u}_{3,k,k} t_{3,k,k}) \\
             &\rev{\mybar{t}_4} \vgad(t_4) \\
             &t_5 \\
             &(\rev{\mybar{t}_{6,1,1}} \mybar{u}_{6,1,1} t_{6,1,1} \cdots \rev{\mybar{t}_{6,1,k}} \mybar{u}_{6,1,k} t_{6,1,k})
             \cdots
             (\rev{\mybar{t}_{6,k,1}} \mybar{u}_{6,k,1} t_{6,k,1} \cdots \rev{\mybar{t}_{6,k,k}} \mybar{u}_{6,k,k} t_{6,k,k}) \\
             &(\rev{\mybar{t}_{7,1,1}} \mybar{u}_{7,1,1} t_{7,1,1} \cdots \rev{\mybar{t}_{7,1,k}} \mybar{u}_{7,1,k} t_{7,1,k})
             \cdots
             (\rev{\mybar{t}_{7,k,1}} \mybar{u}_{7,k,1} t_{7,k,1} \cdots \rev{\mybar{t}_{7,k,k}} \mybar{u}_{7,k,k} t_{7,k,k}) \\
             &\rev{\mybar{t}_8}
    \end{aligned}
    \]
    for some $k$-cliques $t_1, t_2, t_{3,i,j}, t_4, t_5, t_{6,i,j}, t_{7,i,j}, t_8$ and vertices $u_{3,i,j}, u_{6,i,j}, u_{7,i,j}$, where $i,j \in \intv{1}{k}$. 
    By \cref{lem:mu}(b) and because $\mu(s) = \empstr$, we have $t_2 = t_{3,i,j} = t_4$ and $t_5 = t_{6,i,j} = t_{7,i,j} = t_8$ for all $i,j \in \intv{1}{k}$.  We also have
    \[
        \mu\Big(\vgad(t_1) \vgad(t_1) \mybar{u}_{3,1,1} \cdots \mybar{u}_{3,k,k} \vgad(t_2) \mybar{u}_{6,1,1} \cdots \mybar{u}_{6,k,k} \mybar{u}_{7,1,1} \cdots \mybar{u}_{7,k,k} \Big) = \empstr.
    \]
    Applying \cref{lem:mu}(b) again, every vertex of $t_1$ is adjacent to every vertex of $t_2$.  Similarly, $t_2$ and $t_5$ are fully connected.  Then
    \[
        \mu\Big(\vgad(t_1) \mybar{u}_{7,1,1} \cdots \mybar{u}_{7,k,k} \Big) = \empstr
    \]
    and thus $t_1$ and $t_5$ are also fully connected.  Therefore, $t_1 \cup t_2 \cup t_5$ is a $3k$-clique of $G$.
\end{proof}

\subsection{Proofs of \texorpdfstring{\cref{lem:cfl-ds}}{Lemma \getrefnumber{lem:cfl-ds}} and \texorpdfstring{\cref{cor:cfl-ds}}{Lemma \getrefnumber{cor:cfl-ds}}}
\label[appendix]{app:pf-cfl-ds}

\begin{proof}[Proof of \cref{lem:cfl-ds}]
    We largely follow the argument in the textbook \cite[Theorem~10.5.1]{Harrison1978}.
    Let $\Sigma$ be an alphabet and $L \subseteq \Sigma^*$ a context-free language.  We construct a reduction from $L$ to Dyck Selection with $\blu = 1$.  Because the empty input can be handled separately, we may assume that $\empstr \notin L$.  
   
    Fix a context-free grammar $G$ generating $L$.  We may assume that $G$ is in Greibach normal form, that is, every production rule is of the form $A \to a \xi$ where $\xi$ is a (possibly empty) sequence of nonterminals not containing the start symbol $S$.  
Enumerate the nonterminals of $G$ as $A_1, \dots, A_m$ with $A_1 = S$.
For each production rule $p \colon A_i \to a A_{j_1} \cdots A_{j_\ell}$, let $\tau(p) = \rdyck{1} \rdyck{2}^i \rdyck{1} \ldyck{1} \ldyck{2}^{j_{\ell}} \ldyck{1} \cdots \ldyck{1} \ldyck{2}^{j_1} \ldyck{1}$.  
Given a nonempty string $w \in \Sigma^*$, the reduction outputs a Dyck Selection instance $\mathcal{S}_w = S_0, \dots, S_{|w|}$ with $S_0 = \{ \ldyck{1} \ldyck{2} \ldyck{1} \}$ and $S_i = \{ \tau(p) \mid p \in P_{w[i]} \}$ for each $1 \le i \le |w|$, where $P_{w[i]}$ is the set of production rules whose right-hand side starts with $w[i]$.  This can be done in constant space.

    It remains to show that $w \in L$ if and only if $\mathcal{S}_w$ is a yes-instance.
    Let $Q$ be the set of all (possibly empty) sequences of nonterminals not containing $S$, and for each $\xi = A_{i_1} \cdots A_{i_\ell} \in Q$, define $\myenc(\xi) = \ldyck{1} \ldyck{2}^{i_\ell} \ldyck{1} \cdots \ldyck{1} \ldyck{2}^{i_1} \ldyck{1}$.  We write $\derives$ for the one-step derivation relation of $G$ and $\derives^*$ for its reflexive transitive closure.
    \begin{claim} \label{clm:cfl-ds-key}
        Let $k \ge 0$ and $a_1, \dots, a_k \in \Sigma$, and let $\xi \in Q$.  Then $S \derives^* a_1 \cdots a_k \xi$ if and only if there exist $s_i \in S_i$ for $i \in \intv{0}{k}$ such that $\mu(s_0 \cdots s_k) = \myenc(\xi)$.
    \end{claim}
    \begin{proof}
        We proceed by induction on $k$.  The base case $k = 0$ is clear.  For $k \ge 1$, we have
        \begin{align*}
                 & S \derives^* a_1 \cdots a_k \xi \\
            \iff & \exists\,A_i \to a_k \xi_i \in P_{a_k}, \exists \xi_j \in Q \colon \\
                 & \qquad S \derives^* a_1 \cdots a_{k-1} A_i \xi_j \text{ and } \xi = \xi_i \xi_j \\
            \iff & \exists\,A_i \to a_k \xi_i \in P_{a_k}, \exists \xi_j \in Q, \exists s_l \in S_l \text{ for } l \in \intv{0}{k-1} \colon \\
                 & \qquad \mu(s_0 \cdots s_{k-1}) = \myenc(A_i \xi_j) \text{ and } \xi = \xi_i \xi_j \\
            \iff & \exists s_l \in S_l \text{ for } l \in \intv{0}{k} \colon \mu(s_0 \cdots s_k) = \myenc(\xi),
        \end{align*}
        as required.
    \end{proof}
    Setting $k = n$ and $\xi = \empstr$ completes the proof.
\end{proof}

Next, we prove \cref{cor:cfl-ds}.  
We denote by $\mu_1\colon (\Gamma \cup \mybar{\Gamma})^* \to (\Gamma \cup \mybar{\Gamma})^*$ the function $\mu$ for $\blu = 1$.
Let $\vectostr\colon (\vecset_\blu(\Gamma) \cup \vecset_\blu(\mybar{\Gamma}))^* \to (\Gamma^\blu \cup \mybar{\Gamma}^\blu)^*$ be the homomorphism defined in the proof of \cref{lem:ods-semiere}.
\begin{lemma} 
    For every $w \in (\vecset_\blu(\Gamma) \cup \vecset_\blu(\mybar{\Gamma}))^*$, $\mu_1(\vectostr(\mu(w))) = \mu_1(\vectostr(w))$.
    In particular, $\mu_1(\vectostr(w)) = \empstr$ if $\mu(w) = \empstr$.
\end{lemma}
\begin{proof}
    We proceed by induction on $|w|$.  The base case $w = \empstr$ is immediate.
    If $w = w' x$ for some $x \in \vecset_\blu(\Gamma)$, then
    \[
        \mu_1(\vectostr(\mu (w) )) 
        = \mu_1(\vectostr(\mu (w') x ))
        = \mu_1(\vectostr(\mu (w'))) \vectostr(x)
        = \mu_1(\vectostr(w')) \vectostr(x)
        = \mu_1(\vectostr(w'x))
        = \mu_1(\vectostr(w)).
    \]
    Suppose $w = w' \mybar{y}$ for some $y \in \vecset(\Gamma)$.  We distinguish two cases.
    If $\mu(w') = w'' x$ for some $x \in \vecset(\Gamma)$ with $x R y$, then $\mu(w) = w''$. So 
    \begin{align*}
        \mu_1(\vectostr(w))
        & = \mu_1(\vectostr(w')\vectostr(\mybar{y}))
        = \mu_1(\mu_1(\vectostr(w'))\vectostr(\mybar{y}))
        = \mu_1(\mu_1(\vectostr(\mu(w')))\vectostr(\mybar{y})) \\
        & = \mu_1(\vectostr(w'' x)\vectostr(\mybar{y}))
        = \mu_1(\vectostr(w''))
        = \mu_1(\vectostr(\mu(w))).
    \end{align*}
    Otherwise, we have $\mu(w) = \mu(w') \mybar{y}$.  So
    \begin{align*}
        \mu_1(\vectostr(\mu(w)))
        & = \mu_1(\vectostr(\mu(w'))\vectostr(\mybar{y}))
        = \mu_1(\mu_1(\vectostr(\mu(w')))\vectostr(\mybar{y})) \\
        & = \mu_1(\mu_1(\vectostr(w'))\vectostr(\mybar{y}))
        = \mu_1(\vectostr(w')\vectostr(\mybar{y}))
        = \mu_1(\vectostr(w)).
    \end{align*}
    This completes the proof.
\end{proof}

\begin{lemma} 
    For every $w \in (\vecset_\blu(\Gamma) \cup \vecset_\blu(\mybar{\Gamma}))^*$ and $v \in \vecset_\blu(\Gamma)^*$, if $\mu_1(\vectostr(w)) = \vectostr(v)$, then $\mu(w) = v$.
    In particular, $\mu(w) = \empstr$ if $\mu_1(\vectostr(w)) = \empstr$.
\end{lemma}
\begin{proof}
    We proceed by induction on $|w|$.  The base case $w = \empstr$ is immediate.
    If $w = w' x$ for some $x \in \vecset_\blu(\Gamma)$, we have $\mu_1(\vectostr(w)) = \mu_1(\vectostr(w'))\vectostr(x) = \vectostr(v)$.  By the definition of $\vectostr$, there is $v'$ such that $v' x = v$ and $\vectostr(v') = \mu_1(\vectostr(w'))$.  By the induction hypothesis, we obtain $\mu(w') = v'$, and hence $\mu(w) = \mu(w')x = v'x = v$.

    If $w = w' \mybar{y}$ for some $y \in \vecset(\Gamma)$, we have
        $\mu_1(\vectostr(w))
        = \mu_1(\vectostr(w')\vectostr(\mybar{y}))
        = \vectostr(v)$.
        By repeatedly applying \cref{lem:mu-harrison}\ref{lem:mu-harrison-rewind}, we have $\mu_1(\vectostr(w')) = \vectostr(v)\vectostr(x)$ for some $x \in \vecset_\blu(\Gamma)$ with $x R y$.  By the induction hypothesis, we obtain $\mu(w') = vx$, and hence $\mu(w) = \mu(w' \mybar{y}) = \mu(vx \mybar{y}) = \mu(v) = v$.
        This completes the proof.
\end{proof}

\begin{lemma}
    $L(G_R) = \mu_1^{-1}(\empstr)$.
\end{lemma}
\begin{proof}
    By a straightforward induction using \cref{lem:inv}.
\end{proof}

\begin{corollary} \label{cor:gr}
    For every $w \in (\vecset_\blu(\Gamma) \cup \vecset_\blu(\mybar{\Gamma}))^*$, $w \in \dyckset_R$ if and only if $\vectostr(w) \in L(G_R)$.
\end{corollary}
\begin{proof}
    Immediate from the three lemmas above and \cref{lem:mudyck}.
\end{proof}

\begin{proof}[Proof of \cref{cor:cfl-ds}]
We prove that GDS with restriction set $\dyckset_1$ is in LOGCFL by reducing it to a context-free language.
Let $G_R$ be the context-free grammar with the single nonterminal $S$ and production rules $S \to \empstr$ and $S \to S a S \mybar{b}$ for every $a,b \in \Gamma$ satisfying $a R b$, and $L(G_R)$ be the language generated by $G_R$.  Given a GDS instance $\mathcal{S} = S_1, \dots, S_n$, we encode it as 
    \[
        w_{\mathcal{S}} = \bigodot_{i=1}^{n} \left( \left( \bigodot_{s \in S_i} \hash \vectostr(s) \right) \hash \dollar \right).
    \]
    The string $w_{\mathcal{S}}$ can be computed in $\asympO(\log{\blu})$ space.
We then define the following language $L_{\mathcal{S}}$ over $\Sigma = \Gamma \cup \mybar{\Gamma} \cup \{ \hash, \dollar \}$:
\[
    L_{\mathcal{S}} = \left\{ u_1 \hash \vectostr(t_1) \hash v_1 \dollar \cdots u_\ell \hash \vectostr(t_\ell) \hash v_\ell \dollar \;\middle|\;
        \begin{aligned}
            & \ell \ge 0, \\
            & u_i, v_i \in (\Sigma \setminus \{\dollar\})^*, \\
            & t_i \in (\vecset_\blu(\Gamma) \cup \vecset_\blu(\mybar{\Gamma}))^*,  \\
            & \vectostr(t_1) \cdots \vectostr(t_\ell) \in L(G_R)
        \end{aligned}
    \right\}.
\]
A pushdown automaton for $L_{\mathcal{S}}$ can be constructed from one for $L(G_R)$ using $\asympO(\log{\blu})$ space to generate a sequence of $\blu$ states. Hence, $L_{\mathcal{S}}$ is context-free.
By \cref{cor:gr}, we have
\[
    \mathcal{S} \text{ is a yes-instance} \iff \exists s_i \in S_i \text{ for } i \in \intv{1}{n}\colon \vectostr(s_1) \cdots \vectostr(s_n) \in L(G_R) \iff w_{\mathcal{S}} \in L_{\mathcal{S}},
\]
and therefore GDS with restriction set $\dyckset_1$ is log-space reducible to $L_{\mathcal{S}}$.
\end{proof}

\subsection{Proof of \texorpdfstring{\cref{lem:ods-semiere}}{Lemma \getrefnumber{lem:ods-semiere}}}
\label[appendix]{app:pf-ods-semiere}

\begin{lemma} \label{lem:decomp-aux}
    \begin{enumerate}
        \renewcommand{\labelenumi}{\textup{(\alph{enumi})}}
        \item Every $w \in \dyckset_R$ can be decomposed as $w = x_1 w_1 \mybar{y}_1 \cdots x_\ell w_\ell \mybar{y}_\ell$ with $\ell \ge 0$, $x_j, y_j \in \vecset(\Gamma)$ satisfying $x_j R y_j$ and $w_j \in \dyckset_R$ for $j \in \intv{1}{\ell}$.
        \item If $\ldyck{1} w_1 \rdyck{1} \cdots \ldyck{1} w_\ell \rdyck{1} = \ldyck{1} w'_1 \rdyck{1} \cdots \ldyck{1} w'_{\ell'} \rdyck{1}$ for $\ell, \ell' \ge 0$ and $w_1, \dots, w_\ell, w'_1, \dots, w'_{\ell'} \in \dyckset_1$, then $\ell = \ell'$ and $w_1 = w'_1, \dots, w_\ell = w'_\ell$.
    \end{enumerate}
\end{lemma}
\begin{proof}
    A straightforward induction on $|w|$, using \cref{lem:inv}, proves (a).
    For (b), we proceed by induction on $\ell$.  The base case $\ell = 0$ is trivial.
    Suppose $\ell \ge 1$ (and hence $\ell' \ge 1$).  If $w_1 = w'_1$, the claim follows immediately from the induction hypothesis.  Otherwise, assume without loss of generality that $w_1 \rdyck{1}$ is a prefix of $w'_1$.
    By \cref{lem:mu-harrison}, $\mu(w_1 \rdyck{1}) = \mu(\mu(w_1) \rdyck{1}) = \mu(\rdyck{1}) = \rdyck{1} \in \Gamma_1^*$, a contradiction. 
\end{proof}

\begin{corollary} \label{cor:decomp-aux-dyck1}
    Let $\dep \ge 1$.
    Every $w \in \dyckset_1$ can be uniquely decomposed as $w = \ldyck{1} w_1 \rdyck{1} \cdots \ldyck{1} w_\ell \rdyck{1}$ with $\ell \ge 0$ and $w_1, \dots, w_\ell \in \dyckset_1$.  Moreover, if $w \in \dyckset_1^{(\dep)}$, then $w_1, \dots, w_\ell \in \dyckset_1^{(\dep-1)}$.
\end{corollary}

\begin{lemma}[decomposition] \label{lem:decomp}
    Let $\dep \ge 1$.
    Every $w \in \bq^{-1}(\dyckset_1^{(\dep)})\cap \dyckset_R$ can be decomposed as $w = x_1 w_1 \mybar{y}_1 \cdots x_\ell w_\ell \mybar{y}_\ell$ with $\ell \ge 0$, $x_j, y_j \in \vecset(\Gamma)$ satisfying $x_j R y_j$ and $w_j \in \bq^{-1}(\dyckset_1^{(\dep-1)}) \cap \dyckset_R$ for $j \in \intv{1}{\ell}$.
\end{lemma}

\begin{proof}
Let $w = x_1 w_1 \mybar{y}_1 \cdots x_\ell w_\ell \mybar{y}_\ell$ be a decomposition given by \cref{lem:decomp-aux}(a).
Applying $\bq$, we obtain $\bq(w) = \ldyck{1} \bq(w_1) \rdyck{1} \cdots \ldyck{1} \bq(w_\ell) \rdyck{1}$.  By \cref{cor:indyck1}, this gives the decomposition of $\bq(w)$ in \cref{cor:decomp-aux-dyck1}.  Thus, if $\bq(w) \in \dyckset_1^{(\dep)}$, then $\bq(w_1), \dots, \bq(w_\ell) \in \dyckset_1^{(\dep-1)}$, as required.
\end{proof}

\begin{proof}[Proof of \cref{lem:ods-semiere}]
    We prove that the given Generalized Dyck Selection instance $S_1, \dots, S_n$ is a yes-instance if and only if $r$ matches $w$.
    We first prove the following two characterizations.
    \begin{claim} \label{clm:ods-semiere-rzero}
        Let $w' \in \Sigma^*$ and $r'$ a semi-ERE.
        For every $j \in \intv{0}{\blu}$, $\rzeroi{j}\dia{r'}$ matches $w'$ if and only if 
        $w' = \vectostr(x) u \vectostr(\mybar{y})$ for some $u \in r'$ and $x,y \in \vecset_j(\Gamma)$ with $x R y$.
    \end{claim}
    \begin{proof}
        We proceed by induction on $j$.  The base case $j = 0$ is immediate.
        For $j \ge 1$, we have
        \begin{align*}
                 & w' \in \rzeroi{j}\dia{r'} \\
            \iff & \exists a,b \in \Gamma \text{ with } a R b \colon w' \in a \rzeroi{j-1}\dia{r'} \mybar{b} \\
            \iff & \exists u \in r',\,\exists x',y' \in \vecset_{j-1}(\Gamma) \text{ with } x' R y',\,\exists a,b \in \Gamma \text{ with } aRb\colon \\
                 & \qquad w' = a \vectostr(x') u \vectostr(\mybar{y}') \mybar{b} \\
            \iff & \exists u \in r',\,\exists x,y \in \vecset_{j}(\Gamma) \text{ with } x R y\colon w' = \vectostr(x) u \vectostr(\mybar{y}),
        \end{align*}
        completing the proof.
    \end{proof}
    For $w' \in \Sigma^*$ and $v = v_1 \cdots v_\ell \in (\vecset(\Gamma) \cup \vecset(\mybar{\Gamma}))^*$ $(\ell \ge 0, v_1, \dots, v_\ell \in \vecset(\Gamma) \cup \vecset(\mybar{\Gamma}))$, we write $w' \supseq v$ (or $v \subseq w'$) if $w' \in r_0 \vectostr(v_1) \cdots r_0 \vectostr(v_\ell) r_0$.
    \begin{claim} \label{clm:ods-semiere-ri}
        Let $w' \in \Sigma^*$.
        For every $i \in \intv{0}{\dep}$, $r_i$ matches $w'$ if and only if
        there exists $v \subseq w'$ such that $v \in \bq_\blu^{-1}(\dyckset_1^{(i)}) \cap \dyckset_R$.
    \end{claim}
    \begin{proof}
        We proceed by induction on $i$.  The base case $i = 0$ clearly holds.  For $i \ge 1$, by applying \cref{lem:decomp}, the induction hypothesis and \cref{clm:ods-semiere-rzero} in this order, we have 
        \begin{align*}
                 & \exists v \in \bq_\blu^{-1}(\dyckset_1^{(i)}) \cap \dyckset_R \colon w' \supseq v \\
            \iff & \exists \ell \ge 0, \ \exists x_j, y_j \in \vecset_\blu(\Gamma) \text{ with } x_j R y_j, \ \exists v_j \in \bq_\blu^{-1}(\dyckset_1^{(i-1)}) \cap \dyckset_R \text{ for } j \in \intv{1}{\ell} \colon \\
                 & \qquad w' \supseq x_1 v_1 \mybar{y}_1 \, x_2 v_2 \mybar{y}_2 \cdots x_\ell v_\ell \mybar{y}_\ell \\
            \iff & \exists \ell \ge 0, \ \exists x_j, y_j \in \vecset_\blu(\Gamma) \text{ with } x_j R y_j,\ \exists u_j \in r_{i-1},\ \exists z_1, \dots, z_{\ell+1} \in r_0 \text{ for } j \in \intv{1}{\ell} \colon \\
                 & \qquad w' = z_1 \, \vectostr(x_1) u_1 \vectostr(\mybar{y}_1) \, z_2 \, \vectostr(x_2) u_2 \vectostr(\mybar{y}_2) \cdots z_\ell \, \vectostr(x_\ell) u_\ell \vectostr(\mybar{y}_\ell) \, z_{\ell+1} \\
            \iff & w' \in r_i,
        \end{align*}
        completing the proof.
    \end{proof}
    The desired equivalence now follows from \cref{clm:ods-semiere-ri} and the construction of $w$.
\end{proof}

\subsection{Proof of \texorpdfstring{\cref{lem:ods-rsq}}{Lemma \getrefnumber{lem:ods-rsq}}}
\label[appendix]{app:pf-ods-rsq}

\begin{proof}[Proof of \cref{lem:ods-rsq}]
    We prove that the given Orthogonal Dyck Selection instance $S_1, \dots, S_n$ is a yes-instance if and only if $r$ matches $w$.  We call a string $w' \in \Sigma^*$ \emph{good} if it contains no string in $\cent_1 1^* \mybar{\cent}_1$ or $\mybar{\cent}_2 1^* \cent_2$ as a substring.  Observe that $w$ and all its substrings are good.
    \begin{claim} \label{clm:ods-rsq-rzero}
        Let $w' \in \Sigma^*$ be good and $r'$ an RSQ.
        For every $j \in \intv{0}{\blu}$, $\rzeroi{j}\dia{r'}$ matches $w'$ if and only if 
        $w' = \myenc(\vectostr(x))\,u\,\myenc(\vectostr(\mybar{y}))$ for some $u \in r'$ and orthogonal $x,y \in \vecset_j(\Gamma)$.
    \end{claim}
    \begin{proof}
        We proceed by induction on $j$.  The base case $j = 0$ is clear.
        For $j \ge 1$, we have
        \begin{align*}
                 & w' \in \rzeroi{j}\dia{r'} \\
            \iff & w' \in \cent_1 ( 1\cent_2 \,\rzeroi{j-1}\dia{r'}\, \mybar{\cent}_2 1 \regor 1\cent_2 \,\rzeroi{j-1}\dia{r'}\, \mybar{\cent}_2 11 \regor 11\cent_2 \,\rzeroi{j-1}\dia{r'}\, \mybar{\cent}_2 1 ) \mybar{\cent}_1 \\
            \iff & w' \in \myenc(a)\,\rzeroi{j-1}\dia{r'}\,\myenc(\mybar{b}) \text{ for some } a, b \in \{0,1\} \text{ with } a \cdot b = 0 \\
            \iff & w' = \myenc(\vectostr(x)) \,u\, \myenc(\vectostr(\mybar{y})) \text{ for some } u \in r' \text{ and orthogonal } x, y \in \vecset_j(\Gamma),
        \end{align*}
        where the first and third equivalences use the fact that $w'$ and all its substrings are good.
    \end{proof}
    For $w' \in \Sigma^*$ and $v = v_1 \cdots v_\ell \in (\vecset(\Gamma) \cup \vecset(\mybar{\Gamma}))^*$ $(\ell \ge 0, v_1, \dots, v_\ell \in \vecset(\Gamma) \cup \vecset(\mybar{\Gamma}))$, we write $w' \supseq_{\myenc} v$ (or $v \subseq_{\myenc} w'$) if $w' \in r_0\,\myenc(\vectostr(v_1)) \cdots r_0\,\myenc(\vectostr(v_\ell))\,r_0$.
    \begin{claim} \label{clm:ods-rsq-ri}
        Let $w' \in \Sigma^*$ be good.
        For every $i \in \intv{0}{\dep}$, $r_i$ matches $w'$ if and only if
        there exists $v \subseq_{\myenc} w'$ such that $v \in \bq_\blu^{-1}(\dyckset_1^{(i)}) \cap \disjset_2$.
    \end{claim}
    \begin{proof}
        The proof is similar to that of \cref{clm:ods-semiere-ri}.
        We proceed by induction on $i$.  The base case $i = 0$ clearly holds.  For $i \ge 1$, by applying \cref{lem:decomp}, the induction hypothesis and \cref{clm:ods-rsq-rzero} in this order, we have
        \begin{align*}
                 & \exists v \in \bq_\blu^{-1}(\dyckset_1^{(i)}) \cap \disjset_2 \colon w' \supseq_{\myenc} v \\
            \iff & \exists \ell \ge 0, \ \exists\,\text{orthogonal } x_j, y_j \in \vecset_\blu(\Gamma), \ \exists v_j \in \bq_\blu^{-1}(\dyckset_1^{(i-1)}) \cap \disjset_2 \text{ for } j \in \intv{1}{\ell} \colon \\
                 & \qquad w' \supseq_{\myenc} x_1 v_1 \mybar{y}_1 \, x_2 v_2 \mybar{y}_2 \cdots x_\ell v_\ell \mybar{y}_\ell \\
            \iff & \exists \ell \ge 0, \ \exists\,\text{orthogonal } x_j, y_j \in \vecset_\blu(\Gamma),\ \exists u_j \in r_{i-1},\ \exists z_1, \dots, z_{\ell+1} \in r_0 \text{ for } j \in \intv{1}{\ell} \colon \\
                 & \qquad w' = z_1 \, \myenc(\vectostr(x_1)) \, u_1 \, \myenc(\vectostr(\mybar{y}_1)) \, z_2 \cdots z_\ell \, \myenc(\vectostr(x_\ell)) \, u_\ell \, \myenc(\vectostr(\mybar{y}_\ell)) \, z_{\ell+1} \\
            \iff & w' \in r_i,
        \end{align*}
        completing the proof.
    \end{proof}
    The desired equivalence now follows from \cref{clm:ods-rsq-ri} and the construction of $w$.
\end{proof}

\subsection{Proof of \texorpdfstring{\cref{lem:ods-ere}}{Lemma \getrefnumber{lem:ods-ere}}}
\label[appendix]{app:pf-ods-ere}

We claim the following extension of the decomposition lemma (\cref{lem:decomp}), which keeps track of the parameter $\dg$ bounding the degree.  The proof is analogous and is omitted.

\begin{lemma} \label{lem:decomp-deg}
    Let $\dep,\dg \ge 1$.
    Every $w \in \bq^{-1}(\dyckset_1^{\depdg{\dep}{\dg}}) \cap \dyckset_R$ can be decomposed as $w = x_1 w_1 \mybar{y}_1 \cdots x_\ell w_\ell \mybar{y}_\ell$ with $\ell \in \intv{0}{\dg}$, $x_j R y_j$ and $w_j \in \bq^{-1}(\dyckset_1^{\depdg{\dep-1}{\dg}}) \cap \dyckset_R$ for $j \in \intv{1}{\ell}$.
\end{lemma}

\begin{proof}[Proof of \cref{lem:ods-ere}]
    We prove that the given Generalized Dyck Selection instance $S_1, \dots, S_n$ is a yes-instance if and only if $r$ matches $w$.
    For $i \in \intv{0}{h}$, we call a string $w' \in \Sigma^*$ \emph{$i$-good} if it can be written as $w' = \myenc_i(\vectostr(x)) \,u\, \myenc_i(\vectostr(\mybar{y}))$ for some $u \in \Sigma^*$ and $x,y \in \vecset_{\ld^{(h-i)}(\blu)}(\Gamma)$, in which case we write $\leftpart(w') = x$, $\midpart(w') = u$ and $\rightpart(w') = y$, and, for $j,k\in\intv{1}{\ld^{(h-i)}(\blu)}$, let $w'\encintv{j}{k}$ denote the substring
    \[
        x_j\ \bindelim{\cent}{1,i}x_{j+1}\ \cdots\ \bindelim{\cent}{1,i}x_{\ld^{(h-i)}(\blu)} \,\bindelim{\percent}{i}\, u \,\bindelimc{\percent}{i} \, \rev{\mybar{y}}_{\ld^{(h-i)}(\blu)}\bindelimc{\cent}{1,i}\ \cdots\ \rev{\mybar{y}}_{k+1}\bindelimc{\cent}{1,i}\ \rev{\mybar{y}}_k,
    \]
    where
    $z_l = \myenc_{i-1}(\mybin_{\ld^{(h-i+1)}(\blu)}(l))\,\bindelim{\cent}{2,i}\,z[l]$
    for $z \in \{x,y\}$ and $l \in \intv{1}{\ld^{(h-i)}(\blu)}$.
    Observe that $w'\encintv{j}{k}$ is $(i-1)$-good for all $j, k$.

    \begin{claim} \label{clm:ods-ere-req}
        Let $i \in \intv{0}{h-1}$.  Let $w' \in \Sigma^*$ be $i$-good.
        Then $\reqi{\ld^{(h-i)}(\blu)}$ matches $w'$ if and only if $\leftpart(w') = \rightpart(w')$.
    \end{claim}
    \begin{proof}
        We proceed by induction on $i$.  The base case $i = 0$ is clear.  
        For $i \ge 1$, we have
        \begin{align*}
                 & w' \in \reqi{\ld^{(h-i)}(\blu)} \\
            \iff & \forall j,k \in \intv{1}{\ld^{(h-i)}(\blu)}\colon \\
                 & \qquad w'\encintv{j}{k} \notin \reqi{\ld^{(h-i+1)}(\blu)} \regand \rfree{\bindelim{\cent}{2,i}}\,\bindelim{\cent}{2,i}\, \regnot \req\,\bindelimc{\cent}{2,i}\,\rfree{\bindelimc{\cent}{2,i}} \\
            \iff & \forall j,k \in \intv{1}{\ld^{(h-i)}(\blu)}\colon 
                   \mybin_{\ld^{(h-i+1)}(\blu)}(j) = \mybin_{\ld^{(h-i+1)}(\blu)}(k) \Rightarrow \\
                 & \qquad w'\encintv{j}{k} \notin \rfree{\bindelim{\cent}{2,i}}\,\bindelim{\cent}{2,i}\, \regnot\req\,\bindelimc{\cent}{2,i}\,\rfree{\bindelimc{\cent}{2,i}} \\
            \iff & \forall j \in \intv{1}{\ld^{(h-i)}(\blu)}\colon \leftpart(w')[j] = \rightpart(w')[j] \\
            \iff & \leftpart(w') = \rightpart(w'),
        \end{align*}
        completing the proof.
    \end{proof}

    For simplicity, we write $\myenc(-)$ for $\myenc_h(-)$.

    \begin{claim} \label{clm:ods-ere-rzero}
        Let $w' \in \Sigma^*$ be $h$-good and $r'$ a star-free ERE.
        Then $\rzeroi{\blu}\dia{r'}$ matches $w'$ if and only if
        $r'$ matches $\midpart(w')$ and $\leftpart(w')\,R\,\rightpart(w')$.
    \end{claim}
    \begin{proof}
        By \cref{clm:ods-ere-req}, we have
        \begin{align*}
                 & w' \in \rzeroi{\blu}\dia{r'} \\
            \iff & 
               \midpart(w') \in r' \text{ and } 
                   \forall i,j \in \intv{1}{\blu}\colon w'\encintv{i}{j} \notin \reqi{\ld(\blu)} \regand \rfree{\bindelim{\cent}{2,h}}\,\bindelim{\cent}{2,h}\, \regnot r_R\,\bindelimc{\cent}{2,h}\,\rfree{\bindelimc{\cent}{2,h}} 
               \\
            \iff & 
               \midpart(w') \in r' \text{ and } 
               \left(
                   \begin{aligned}
                       & \forall i,j \in \intv{1}{\blu}\colon  
                       \mybin_{\ld(\blu)}(i) = \mybin_{\ld(\blu)}(j) \Rightarrow \\
                       & \qquad w'\encintv{i}{j} \notin \rfree{\bindelim{\cent}{2,h}}\,\bindelim{\cent}{2,h}\, \regnot r_R\,\bindelimc{\cent}{2,h}\,\rfree{\bindelimc{\cent}{2,h}}
                   \end{aligned}
               \right) \\
            \iff & 
               \midpart(w') \in r' \text{ and } 
                   \forall i \in \intv{1}{\blu}\colon \leftpart(w')[i]\, R\, \rightpart(w')[i],
        \end{align*}
        completing the proof.
    \end{proof}
    For $w' \in \Sigma^*$ and $v = v_1 \cdots v_\ell \in (\vecset(\Gamma) \cup \vecset(\mybar{\Gamma}))^*$ $(\ell \ge 0, v_1, \dots, v_\ell \in \vecset(\Gamma) \cup \vecset(\mybar{\Gamma}))$, we write $w' \supseq_{\myenc} v$ (or $v \subseq_{\myenc} w'$) if $w' \in r_0 \,\myenc(\vectostr(v_1)) \cdots r_0\,\myenc(\vectostr(v_\ell))\,r_0$.
    \begin{claim} \label{clm:ods-ere-ri}
        Let $w'$ be a substring of $w$ that is empty or, for some $x_1, x_2 \in \vecset_\blu(\Gamma) \cup \vecset_\blu(\mybar{\Gamma})$, begins with $\myenc(\vectostr(x_1))$ and ends with $\myenc(\vectostr(x_2))$.
        For every $i \in \intv{0}{\dep}$, $r_i$ matches $w'$ if and only if
        there exists $v \subseq_{\myenc} w'$ such that $v \in \bq_\blu^{-1}(\dyckset_1^{\depdg{i}{\dg}}) \cap \dyckset_R$.
    \end{claim}
    \begin{proof}
        We proceed by induction on $i$.  The base case $i = 0$ clearly holds.  
        For $i \ge 1$, by applying \cref{lem:decomp-deg}, the induction hypothesis and \cref{clm:ods-ere-rzero} in this order, we have
        \begin{align*}
                 & \exists v \in \bq_\blu^{-1}(\dyckset_1^{\depdg{i}{\dg}}) \cap \dyckset_R \colon w' \supseq_{\myenc} v \\
            \iff & \exists \ell \in \intv{0}{\dg}, \ \exists x_j, y_j \in \vecset_\blu(\Gamma) \text{ with } x_j R y_j, \ \exists v_j \in \bq_\blu^{-1}(\dyckset_1^{\depdg{i-1}{\dg}}) \cap \dyckset_R \text{ for } j \in \intv{1}{\ell} \colon \\
                 & \qquad w' \supseq_{\myenc} x_1 v_1 \mybar{y}_1 \, x_2 v_2 \mybar{y}_2 \cdots x_\ell v_\ell \mybar{y}_\ell \\
            \iff & \exists \ell \in \intv{0}{\dg}, \ \exists x_j, y_j \in \vecset_\blu(\Gamma) \text{ with } x_j R y_j,\ \exists u_j \in r_{i-1},\ \exists z_1, \dots, z_{\ell+1} \in r_0 \text{ for } j \in \intv{1}{\ell} \colon \\
                 & \qquad w' = z_1 \, \myenc(\vectostr(x_1)) \, u_1 \, \myenc(\vectostr(\mybar{y}_1)) \, z_2 \cdots z_\ell \, \myenc(\vectostr(x_\ell)) \, u_\ell \, \myenc(\vectostr(\mybar{y}_\ell)) \, z_{\ell+1} \\
            \iff & \exists \ell \in \intv{0}{\dg}, \ \exists\,\text{$h$-good } w'_j \in \rzeroi{\blu}\dia{r_{i-1}}, \ \exists z_1, \dots, z_{\ell+1} \in r_0 \text{ for } j \in \intv{1}{\ell} \colon \\
                 & \qquad w' = z_1 w'_1 z_2 w'_2 \cdots z_\ell w'_\ell z_{\ell+1} \\
            \iff & w' \in r_i,
        \end{align*}
        where, in the last $\Longleftarrow$ direction, the $h$-goodness condition follows from the assumption on $w'$.
    \end{proof}
    The desired equivalence now follows from \cref{clm:ods-ere-ri} and the construction of $w$.
\end{proof}

\subsection{Proofs of \texorpdfstring{\cref{lem:kov-krewb}}{Lemma \getrefnumber{lem:kov-krewb}} and \texorpdfstring{\cref{lem:ov-rewb}}{Lemma \getrefnumber{lem:ov-rewb}}}
\label[appendix]{app:pf-ov-rewb}

\begin{proof}[Proof of \cref{lem:kov-krewb}]
We prove that there exist vectors $a_1 \in A_1, \dots, a_{2k} \in A_{2k}$ that are orthogonal if and only if $r$ matches $w$.  For the only if direction, we can decompose $s$ as
    \[
        s = \bigodot_{i=1}^{k} w_{4(i-1)} a_{2i-1} \hash w_{4i-3} \dollar w_{4i-2} \hash \rev{a_{2i}} w_{4i-1} \dollar
    \]
    for some $w_0, \dots, w_{4k-1} \in [01\hash]^*$.  We let $r$ capture $\hash w_{4i-3} \dollar w_{4i-2} \hash$ in the $i$-th capturing group.  By splitting $w$ using the delimiter $\cent$, it suffices to show that for every $l \in \intv{1}{d}$, the subexpression $\rzeroij{l}{i}$ matches $s$ for some $i \in \intv{1}{k}$.  Fix $l \in \intv{1}{d}$.  Because $a_1, \dots, a_{2k}$ are orthogonal, we can choose a vector $a_i$ whose $(d-l+1)$-th coordinate is zero.  It is straightforward to see that $\rzeroij{l}{i/2}$ matches $s$ when $i$ is even, and $\rzeroij{l}{(i+1)/2}$ matches $s$ when $i$ is odd.

    Conversely, suppose that $r$ matches $w$.  The delimiter $\cent$ enforces that the subexpression
    \[
        \bigodot_{i=1}^{k} [01\hash]^* (\hash [01\hash]^* \dollar [01\hash]^* \hash)_i [01\hash]^* \dollar
    \]
    matches $s$ in such a way that for each $i \in \intv{1}{k}$, the $i$-th capturing group matches $\hash w_{4i-3} \dollar w_{4i-2} \hash$ for some $w_{4i-3}, w_{4i-2} \in [01\hash]^*$.  By construction of $w$, there are vectors $a_1 \in A_1, \dots, a_{2k} \in A_{2k}$ such that $s = \bigodot_{i=1}^{k} w_{4i-4} a_{2i-1} \hash w_{4i-3} \dollar w_{4i-2} \hash \rev{a_{2i}} w_{4i-1} \dollar$ for some $w_{4i-4}, w_{4i-1} \in [01\hash]^*$ for $i \in \intv{1}{k}$.  Then, for $l \in \intv{1}{d}$, there is some $i \in \intv{1}{k}$ such that $\rzeroij{l}{i}$ matches $s$ so that $\bs i$ matches $\hash w_{4i-3} \dollar w_{4i-2} \hash$.  The delimiter $\dollar$ enforces that either $a_{2i-1}$ or $a_{2i}$ has zero in its $(d-l+1)$-th coordinate.  By repeating the argument for all $l \in \intv{1}{d}$, we have that $a_1, \dots, a_{2k}$ are orthogonal.
\end{proof}

\begin{proof}[Proof of \cref{lem:ov-rewb}]
We prove that there exist vectors $a_i \in A$ and $b_j \in B$ that are orthogonal if and only if $r$ matches $w$.  For the only if direction, we can decompose $w$ as
    \[
        w = w_{0} a_i \hash w_{1} \dollar w_{2} \hash (\rev{b_j} w_{3} \cent w_{0} a_i \hash w_{1} \dollar w_{2} \hash)^d \rev{b_j} w_{3} \cent
    \]
    where $a_i \in A$ and $b_j \in B$ are orthogonal, and $w_{0}, w_{1}, w_{2}, w_{3} \in [01\hash]^*$.
    We let $r$ capture $\hash w_1 \dollar w_2 \hash$ in its capturing group.  
    By splitting $w$ using the delimiter $\cent$, it suffices to show that $\rev{b_j} w_3 \cent w_0 a_i \in \rzeroi{l}$ for all $l \in \intv{1}{d}$.  This indeed holds because $a_i$ and $b_j$ are orthogonal.
    
    Conversely, suppose that $r$ matches $w$.  The delimiters $\cent$ and $\dollar$ enforce that the subexpression $[01\hash]^* (\hash \Sigma^* \hash)_1$ matches a prefix of $s$ in such a way that the capturing group matches $\hash w_1 \dollar w_2 \hash$ for some $w_1, w_2 \in [01\hash]^*$.  By the construction of $w$, there are vectors $a_i \in A$ and $b_j \in B$ such that $s = w_0 a_i \hash w_1 \dollar w_2 \hash \rev{b_j} w_3$ for some $w_0, w_3 \in [01\hash]^*$.  Then, every subsequent subexpression $\rzeroi{1}, \dots, \rzeroi{d}$ matches $\rev{b_j} w_3 \cent w_0 a_i$.  This implies that $a_i$ and $b_j$ are orthogonal.
\end{proof}

\end{document}